\documentstyle{mn}
\newif\ifAMStwofonts
\def\cm3{cm$^{-3}$}

\def\lsun{L$_{\odot}$}

\ifoldfss
  \ifCUPmtlplainloaded \else
    \NewTextAlphabet{textbfit} {cmbxti10} {}
    \NewTextAlphabet{textbfss} {cmssbx10} {}
    \NewMathAlphabet{mathbfit} {cmbxti10} {} % for math mode
    \NewMathAlphabet{mathbfss} {cmssbx10} {} %  "   "    "
  \fi
  \ifAMStwofonts
    \ifCUPmtlplainloaded \else
      \NewSymbolFont{upmath} {eurm10}
      \NewSymbolFont{AMSa} {msam10}
      \NewMathSymbol{\upi}     {0}{upmath}{19}
      \NewMathSymbol{\umu}     {0}{upmath}{16}
      \NewMathSymbol{\upartial}{0}{upmath}{40}
      \NewMathSymbol{\leqslant}{3}{AMSa}{36}
      \NewMathSymbol{\geqslant}{3}{AMSa}{3E}

      \let\leq=\leqslant \let\le=\leqslant
       \let\ge=\geqslant
    \fi
  \fi
\fi % End of OFSS

\ifnfssone
  \newmathalphabet{\mathit}
  \addtoversion{normal}{\mathit}{cmr}{m}{it}
  \addtoversion{bold}{\mathit}{cmr}{bx}{it}
  \newmathalphabet{\mathbfit} % math mode version of \textbfit{..}
  \addtoversion{normal}{\mathbfit}{cmr}{bx}{it}
  \addtoversion{bold}{\mathbfit}{cmr}{bx}{it}
  \newmathalphabet{\mathbfss} % math mode version of \textbfss{..}
  \addtoversion{normal}{\mathbfss}{cmss}{bx}{n}
  \addtoversion{bold}{\mathbfss}{cmss}{bx}{n}
  \ifAMStwofonts
    \ifCUPmtlplainloaded \else
      \UseAMStwoboldmath
      \makeatletter
      \new@mathgroup\upmath@group
      \define@mathgroup\mv@normal\upmath@group{eur}{m}{n}
      \define@mathgroup\mv@bold\upmath@group{eur}{b}{n}
      \edef\UPM{\hexnumber\upmath@group}
      \new@mathgroup\amsa@group
      \define@mathgroup\mv@normal\amsa@group{msa}{m}{n}
      \define@mathgroup\mv@bold\amsa@group{msa}{m}{n}
      \edef\AMSa{\hexnumber\amsa@group}
      \makeatother
      \mathchardef\upi="0\UPM19
      \mathchardef\umu="0\UPM16
      \mathchardef\upartial="0\UPM40
      \mathchardef\leqslant="3\AMSa36
      \mathchardef\geqslant="3\AMSa3E

      \let\leq=\leqslant \let\le=\leqslant
       \let\ge=\geqslant
    \fi
  \fi
\fi % End of NFSS release 1

\ifnfsstwo
  \DeclareMathAlphabet{\mathbfit}{OT1}{cmr}{bx}{it}
  \SetMathAlphabet\mathbfit{bold}{OT1}{cmr}{bx}{it}
  \DeclareMathAlphabet{\mathbfss}{OT1}{cmss}{bx}{n}
  \SetMathAlphabet\mathbfss{bold}{OT1}{cmss}{bx}{n}
  \ifAMStwofonts
    \ifCUPmtlplainloaded \else
      \DeclareSymbolFont{UPM}{U}{eur}{m}{n}
      \SetSymbolFont{UPM}{bold}{U}{eur}{b}{n}
      \DeclareSymbolFont{AMSa}{U}{msa}{m}{n}
      \DeclareMathSymbol{\upi}{0}{UPM}{"19}
      \DeclareMathSymbol{\umu}{0}{UPM}{"16}
      \DeclareMathSymbol{\upartial}{0}{UPM}{"40}
      \DeclareMathSymbol{\leqslant}{3}{AMSa}{"36}
      \DeclareMathSymbol{\geqslant}{3}{AMSa}{"3E}

      \let\leq=\leqslant \let\le=\leqslant
       \let\ge=\geqslant
    \fi
  \fi
\fi % End of NFSS release 2

\ifCUPmtlplainloaded \else
  \ifAMStwofonts \else % If no AMS fonts
    \def\upi{\pi}
    \def\umu{\mu}
    \def\upartial{\partial}
  \fi
\fi

\title[Properties of hot stars in NGC\,5253 from ISO spectroscopy]
     {Properties of hot stars in the Wolf-Rayet galaxy NGC\,5253 from 
ISO-SWS spectroscopy}
\author[P.A. Crowther et al.]
       {Paul A. Crowther$^1$\thanks{email: pac@star.ucl.ac.uk}, 
S. C. Beck$^2$,  Allan J. Willis$^1$, 
Peter S. Conti$^3$, \cr Patrick W. Morris$^{4,5}$ and Ralph S. Sutherland$^6$ \\
$^1:$ Department of Physics \& Astronomy, University College London,
Gower Street, London WC1E 6BT\\
$^2:$ School of Physics \& Astronomy of the Raymond \& Beverly Sackler Faculty
of Exact Sciences, Tel Aviv University, Ramat Aviv, Israel \\
$^3:$ JILA, University of  Colorado, Boulder, CO 80309, USA \\
$^4:$ Villafranca del Castillo Satellite Tracking Station,  Madrid, Spain\\
$^5:$ SRON, Sorbonnelaan 2, CA 3584 Utrecht, The Netherlands \\
$^6:$ Mount Stromlo and Siding Spring Observatories, 
Australian National University, Canberra, ACT 0200, Australia}
\date{Accepted ???; Received ???}

\pagerange{1-16}
\pubyear{1998}

\begin{document}

\maketitle

\label{firstpage}

\begin{abstract}
ISO-SWS spectroscopy of the WR galaxy NGC\,5253 is presented, and analysed to
provide estimates of its hot young star  population. Our approach differs
from previous investigations in that  we are  able to distinguish
between the regions in which
different infrared fine-structure lines form, using 
complementary ground-based observations. The high excitation  nebular
[S\,{\sc iv}] emission is formed in a very compact region, which we
attribute to the central super-star-nucleus, and lower excitation 
[Ne\,{\sc ii}] nebular emission originates in the galactic core.
We use photo-ionization modelling coupled with the
latest theoretical  O-star flux distributions to derive effective stellar
temperatures and ionization parameters of $T_{\rm eff} \ge 38$kK,
log~$Q \sim 8.25$ for the compact nucleus, with 
$T_{\rm eff} \sim$35kK, $\log~Q \le$8 for the larger core.
Results are supported by more sophisticated calculations using evolutionary 
synthesis models. We assess the contribution that
Wolf-Rayet stars may make to highly ionized nebular lines 
(e.g. [O\,{\sc iv}]).

From our Br$\alpha$ flux, the 2$''$ nucleus contains the equivalent of 
approximately 1\,000 O7\,V star equivalents  and the 
starburst there is 2--3\,Myr old; the 20$''$ core contains 
about 2\,500 O7\,V star equivalents, with a representative age of $\sim$5\,Myr.
The  Lyman ionizing flux of the nucleus is equivalent to 
the 30~Doradus region. These quantities are in good agreement
with the observed mid-IR dust luminosity of 7.8$\times$10$^{8}$ {\lsun}.
Since this structure of hot clusters embedded in cooler emission  may be
common in dwarf starbursts, observing a galaxy solely with a large aperture
may result in confusion. Neglecting the spatial distribution of nebular
emission in NGC\,5253, implies `global' stellar
temperatures (or ages) of 36kK (4.8\,Myr) and 39kK (2.9 or 4.4\,Myr) from
the observed [Ne\,{\sc iii/ii}] and [S\,{\sc iv/iii}] line ratios, assuming 
$\log~Q$=8.
\end{abstract}


\section{Introduction}

Wolf-Rayet (WR) galaxies are a subset of blue emission-line galaxies 
whose spectra show the signature of large numbers of
WR stars (Vacca \& Conti 1992).
These galaxies are among the youngest  starburst galaxies and contain a 
large population of  very massive stars with ages in the range 
1--10$\times$10$^{6}$ yr. Mid infra-red (IR) observations 
are well suited to the study of hot, massive stars since 
this spectral region contains many fine-structure nebular lines which
depend very sensitively on stellar content. Until recently, use of such
diagnostics was severely hindered because of the low transparency of our
atmosphere at mid-IR wavelengths. The launch of the Infrared Space
Observatory (ISO) has opened up this new window, which we are exploiting
through a Guest Observer program (WRGALXS; P.I. A.J.Willis) with the ISO
Short Wavelength Spectrometer (SWS). 

NGC\,5253 is one of the closest WR galaxies, at a distance of only 
4.1\,Mpc (Saha et al. 1995) and represents one of 
the targets of our Guest Observer program. NGC\,5253  contains several very
young super-star clusters, no older than a few million years 
(Walsh \& Roy 1989; Beck et al. 1996). It has  
been  the focus of 
many recent studies (e.g. Gorjian 1996; Calzetti et al. 1997; Turner,
Beck \& Hurt 1997), which have shown that the  
extinction  is very high and patchy at ultraviolet and visual wavelengths.
ISO  observations 
of NGC\,5253 are particularly important since the interstellar extinction 
at mid-IR wavelengths is low and because earlier spectral observations
found evidence for a remarkably hot and young stellar population (Aitken et
al. 1982, Lutz et al. 1996; Beck et al. 1996). 

In Section~\ref{sect2} we present our new ISO-SWS spectroscopy of NGC\,5253 
together with complementary ground-based observations, which are 
interpreted in Section~\ref{sect3}. Our technique
for  photo-ionization modelling, based largely on {\sc cloudy} 
(Ferland 1996) is discussed in Section~\ref{sect4}, 
including assumed elemental abundances and theoretical flux distributions 
for O stars. In Section~\ref{sect5} we discuss the results of our
photo-ionization modelling.
The influence from WN and WC-type Wolf-Rayet  stars is considered in
Section~\ref{sect6}, while results obtained using evolutionary synthesis
models are presented in Section~\ref{sect7}. Our present results are compared 
with those from the literature in Section~\ref{sect8} and our conclusions are 
reached in Section~\ref{sect9}.

\begin{figure}
\vspace{9cm}
\includegraphics{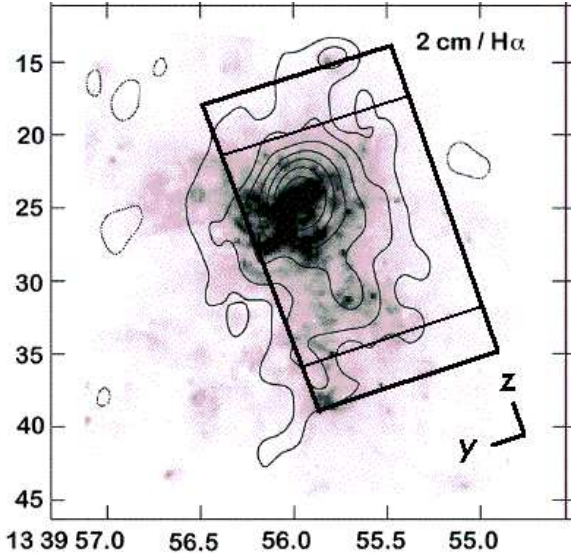}
\caption{ISO--SWS apertures superimposed on the 
Calzetti et al. (1997) HST/WFPC2 H$\alpha$ image
of NGC\,5253, together with the 2 cm VLA radio continuum contour map
from Beck et al. (1996). 
The ISO aperture is 14$\times$20 arcsec for $\lambda$$\le$12$\mu$m,
and 14$\times$27 arcsec at longer wavelengths. The spacecraft Y- and 
Z-axes, indicated here, are discussed in the text. Declination 
($-$31 38m) and right ascension  (13h 39m) are in J2000.0 coordinates.}
\label{fig1a}
\end{figure}

\section{Observations and Reduction}\label{sect2}

% Updated by Pat 12/08/98

Our new infrared data on NGC\,5253 were obtained as part of a Guest 
Observer programme with 
the Short Wavelength Spectrograph (SWS; de Graauw et al. 1996)
onboard the ESA Infrared Space Observatory (Kessler et al. 1996)
in revolution 418, 7 January 1997.  

The SWS AOT6 observing mode was
used to achieve full grating resolution, $\lambda/\Delta\lambda \;
\sim \; 1300-2500$ and the continuous wavelength coverage was 
2.99--27.65\,$\mu$m.  Data of detector ``bands'' 1 and 2 respectively cover 
wavelengths of 2.99--4.08\,$\mu$m using 12 In:Sb detectors and 
4.00--12.05\,$\mu$m with 12 Si:Ga detectors, employing entrance slits
that give an effective aperture area of 14~$\times$~20 arcsec on the sky.  
Band 3 data cover 12.0--27.65\,$\mu$m using 12 Si:As detectors, with sky 
coverage of 14~$\times$~27  arcsec.  The total integration time was set to 
9140 seconds to allow one complete scan over the wavelengths selected 
within each ``AOT band'', defined by the permissible combinations of detector 
band, aperture, and spectral order (cf. de Graauw et al. 1996).  The 
observation time includes dark current measurements, and a monitor of 
photometric drift for the detectors of bands 2--3.  A drift measurement is 
not normally made for the relatively stable In:Sb detectors.

The SWS data were processed through the main stages of raw signal conversion
(bits to $\mu$V/sec), glitch recognition, wavelength calibration, and
dark current subtraction by following the standard chain of pipeline
software and calibration tables complying with version 6.0 or later.  
Final photometric calibrations performed in the SWS Interactive Analysis 
environment rely on relative spectral response functions and absolute
photometric response factors that were conditional for further pipeline 
development.  These are generally consistent with calibrations found
in pipeline version 7.0.
Prior to co-addition of each detector's output, data were statistically
filtered using the results of glitch recognition algorithms, rejecting 
data points whose corresponding signal ramps were found to have more
than two departures from linearity after AC correction.  Detector 27
was judged to behave photometrically unstable in this observation, and
all output from it was rejected.  Remaining detector
output was rectified to the mean flux density within each resolution
elements, 3-$\sigma$ points were excluded, and finally re-binned to
a spectral resolution that is somewhat higher than instrumental. 

General aspects of the wavelength and photometric calibrations as of
revolution $\sim$100 (after initial verification of instrument performance) are
described, respectively, by Valentijn et al. (1996) and Schaeidt et al. (1996).
More contemporary performance details are in preparation for the ISO Post Mission
Archive by the SWS instrument team.

\subsection{Aperture orientation}\label{aperture}

The input coordinates coinciding with aperture centre were 
$\alpha$ = 13:39:55.7, $\delta$ = $-$31:38:29.0 (J2000.0).  Spacecraft 
pointing uncertainties are alleged to be around 2 arcsec at the time of
this pointing, following recalibration of the star-tracker CCD some 
50 revolutions prior. Pointing reconstruction software in pipeline
version 7.0, accounting also for later changes in solar ephemerides updating 
procedures, indicates offsets, both along dispersed (corresponding to 
the spacecraft Z axis) and orthogonal to the dispersed direction
of SWS (Y axis) to be within
the quoted uncertainty.  The centre of our apertures is $\sim$2$''$
from the ultraviolet source UV3  of Kobulnicky et al. (1997), but is sufficiently 
large to include the strong nebular H$\alpha$ and Br$\alpha$  emitting regions 
of NGC\,5253, $\approx$5$''$ from the centre of our aperture (see Kobulnicky et al. 
1997; Davies, Sugai \& Ward 1998). The two SWS aperture projections are 
overlaid on a {\it Hubble Space Telescope} (HST) WFPC2 H$\alpha$ image of 
NGC\,5253 from Calzetti 
et al. (1997) in Fig.~\ref{fig1a}, together with a 2 cm VLA contour map
from Beck et al. (1996).

It is important to note that photometric responses are not flat over the
spatial extent of the area covered by each detector block.  This is most 
important in the direction orthogonal to the dispersion axis (along the 
spacecraft Y axis; 
see Fig.~\ref{fig1a}), where an offset of 5 arcsec from the position 
of peak responsivity results in a loss of 20--25\% flux of combined detector
output from a point source.  The loss within
2 arcsec of the aperture centre 
is less than 5\% for bands 2--3, which exhibit similar
beam profiles.  This is shown in rough form in Schaeidt et al. (1996), but
detailed profiles to aperture edge and beyond are in analysis at this
time. For band 1, the loss is negligible within $\sim$2.5 arcsec of 
the aperture centre, but a sharper 
gradient in responsivity beyond this again results in 
a loss of 20--25\% at $\sim$5 arcsec away.  If 30\% of Br$\alpha$
line emission does originate from the central nucleus of the NGC\,5253 (cf. 
Davies et al. 1998), then the offset
along the minor axis of the aperture may mean that the energy measured
in the line is underestimated by $\sim$10\%.  For more extended line
emission (see below), a loss estimate is more difficult to arrive at without
a surface brightness distribution and knowledge of the photometric
responses to the aperture edges and beyond. Overall, photometric 
uncertainties  are estimated to be $\sim$5\% in band 1, 
10--15\% in band 2, and 15--25\% in band 3. 
No large discrepancies between levels of adjoining 
AOT bands were noticed. 

\begin{figure*}
\vspace{11cm}
\includegraphics{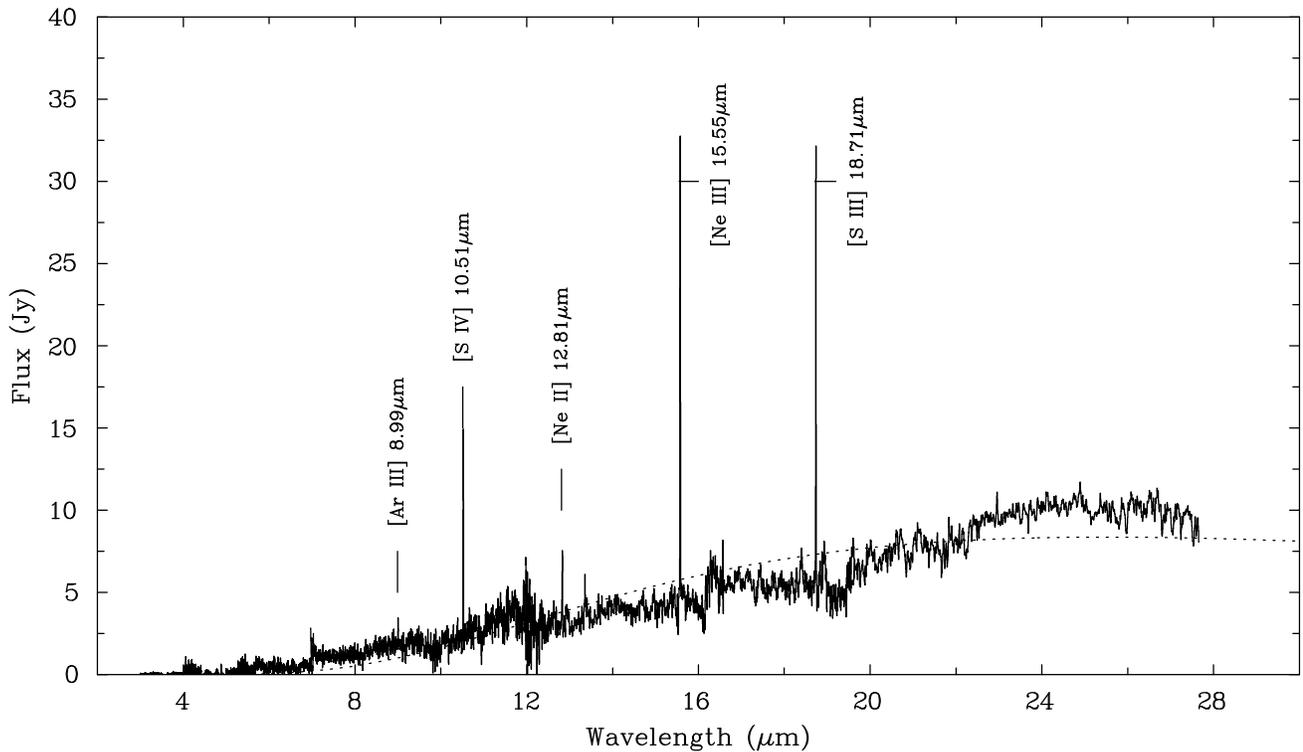}
\caption{Flux calibrated (in Jansky) spectrum of NGC\,5253 
obtained with ISO/SWS, including a black body fit (dotted line) to the 
observed dust continuum. The broad feature between 22--27$\mu$m appears to be
due to solid state silicate emission, representing the first such 
detection in a starburst galaxy.} 
\label{fig1}
\end{figure*}

\subsection{Infrared nebular emission line fluxes}\label{IRfluxes}

We present the full SWS spectrum of NGC\,5253 in Fig.~\ref{fig1}.
Several sharp nebular emission lines are apparent,  superimposed on a 
rising `continuum' flux distribution attributed to dust emission 
(see Section~\ref{dust}).
The most pronounced nebular lines are due to ground--state fine 
structure lines of [S\,{\sc iv}]10.51$\mu$m, [Ne\,{\sc iii}]15.55$\mu$m and 
[S\,{\sc iii}]18.71$\mu$m; weaker nebular emissions are also seen 
in [Ne\,{\sc ii}]12.81$\mu$m and [Ar\,{\sc iii}]8.99$\mu$m. 
We have used the emission line fitting routine {\sc elf} in {\sc dipso}
(Howarth et al. 1995) to measure line fluxes using an eye--estimated 
local  continuum flux, which are presented in Table~\ref{table1}. 
Since the ISO aperture is offset from the central nucleus along the
minor axis, we have included a 10--15\% correction for light loss.
From the Table, there are clearly additional uncertainties -- the measured
[S\,{\sc iv}] flux from the small aperture ground-based observation is
{\it greater} than that from the large aperture ISO data set. 

\begin{figure}
\vspace{12cm}
\includegraphics{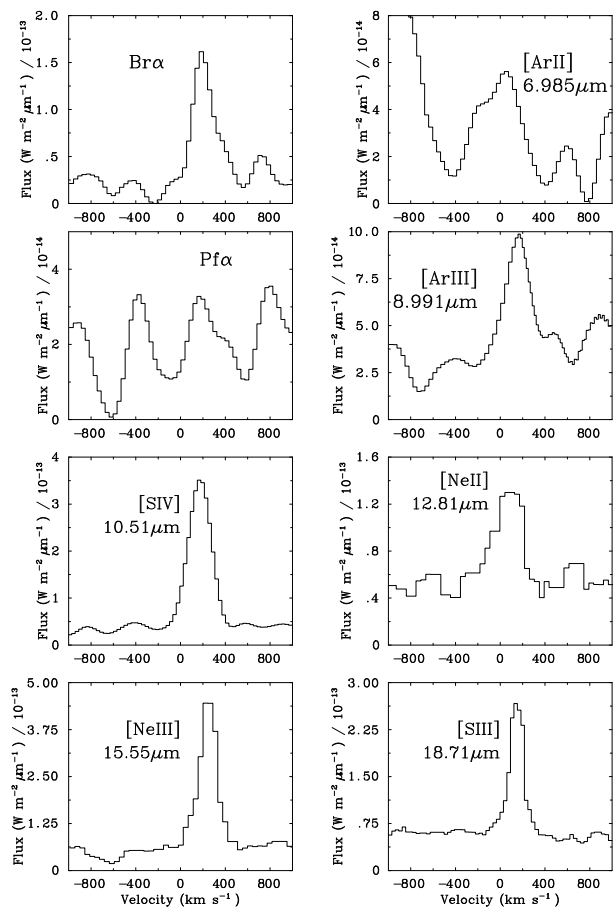}
\caption{Principal nebular emission lines observed in 
the ISO--SWS spectrum of NGC\,5253, red-shifted by 
$\Delta v$$\sim$+200 km s$^{-1}$. The identification of [Ar\,{\sc ii}] and
Pf~$\alpha$ are uncertain.}
\label{fig2}
\end{figure}

Individual nebular line profiles are shown in Fig~\ref{fig2}, although these 
are unresolved at the SWS AOT6 resolving power. Emission lines 
are red-shifted by $\Delta v$$\sim$+200 km s$^{-1}$,
in reasonable agreement with optical measurements of the galactic recession
velocity (+416 km\,s$^{-1}$ from de Valcouleurs et al. 1993). The observed
$\Delta v$ was then used to search for weaker emission lines allowing us to
confirm the presence of Br~$\alpha$~4.051$\mu$m  and (possibly)
Pf$\alpha$~7.458$\mu$m (see  Fig~\ref{fig2}). We include the spectral 
region around [Ar\,{\sc ii}]6.99$\mu$m in our band~2C SWS dataset, although 
the positive identification and measured flux of this feature is uncertain, 
due to its location at the edge of the 2B and 2C bands and low $\Delta v$.

\begin{table} 
\caption[]{Nebular emission lines observed in our ISO--SWS (AOT6)
spectrum of NGC\,5253, and ground--based measurements. Our ISO
fluxes are after correction  for the variable profile across 
the beam in the spacecraft Y-axis (see Sect.~\ref{aperture}).} 
\label{table1} 
\begin{center} 
\begin{tabular}{lr@{\hspace{2mm}}r@{\hspace{4mm}}c@{\hspace{2mm}}l} \hline 
 Ion & $\lambda _{\rm lab}$ & Obs. Flux & Aperture & Source \\ 
 &$\mu$m & 10$^{-16}$W\,m$^{-2}$ & arcsec & \\ 
 \hline 
Br--$\gamma$ & 2.16 &1.5 & 10 $\times$ 20 & Kawara et al. (1989)\\
             & 2.16 &1.0 & 4              & Davies et al. (1998) \\
             & 2.16 &2.9 & 30             & Davies et al. (1998) \\
Br-$\alpha$  & 4.05 &5.1 & 14 $\times$ 20 & ISO--SWS \\ 
             & 4.05 &7.0 & 10 $\times$ 20 & Kawara et al. (1989)\\ 
Ar\,{\sc ii} & 6.99 &$<$4.0 & 14 $\times$ 20 & ISO--SWS (band 2C)\\ 
Pf-$\alpha$  & 7.46 &$\le$1.0 & 14 $\times$ 20 & ISO--SWS \\ 
Ar\,{\sc iii} & 8.99 &5.6 & 14 $\times$ 20 & ISO--SWS \\ 
              & 8.99 &3.6 & 1.6            &Beck et al. (1996)\\
S\,{\sc iv}  & 10.51 &32.4& 14 $\times$ 20 & ISO--SWS \\ 
             & 10.51 &39.0& 1.6            & Beck et al. (1996)\\
Ne\,{\sc ii} & 12.81 &11.5& 14 $\times$ 27 & ISO-SWS \\ 
             & 12.81 &$\leq$1.0& 1.6      & Beck et al. (1996)\\
Ne\,{\sc iii}& 15.55 &46.2& 14 $\times$ 27 & ISO-SWS \\
S\,{\sc iii} & 18.71 &23.0& 14 $\times$ 27 & ISO-SWS \\
S\,{\sc iii} & 33.46 & 35.7& 20 $\times$ 33 & Genzel et al. (1998)\\
Si\,{\sc ii} & 34.81 & 19.5& 20 $\times$ 33 & Genzel et al. (1998)\\
\hline
\end{tabular}
\end{center}
\end{table}

% redshift = 0.001348

We have  also collected IR flux measurements for NGC\,5253 
from the literature which, are also included in 
Table~\ref{table1}. The hydrogen Brackett line data are taken  from 
Kawara, Nishida \& Phillips (1989) and Davies et al. (1998) and the 
metal lines from Beck et al. (1996). 
In general the ISO--SWS and ground-based results agree 
reasonably well, except for the 
[Ne\,{\sc ii}]12.81$\mu$m line (see Section~\ref{sect3}). Genzel et al. (1998)
have recently presented additional [S\,{\sc iii}] and [Si\,{\sc ii}] mid-IR 
line flux ratios obtained with band 4 of ISO/SWS, obtained at a 
location $\approx$2.4 arcsec from our data set. We have converted
these into absolute flux measurements using our [Ne\,{\sc ii}] measurement,
although this introduces an additional uncertainty since the SWS band~4
aperture is 20$\times$33 arcsec. In theory we should be able to use the
observed ratio of [S\,{\sc iii}] (33.5$\mu$m/18.7$\mu$m) of $\sim$1.6
to measure the (emissivity weighted) mean electron density of the 
[S\,{\sc iii}] emitting region. However, these lines were obtained
through different sized apertures, and the observed ratio is close to
the low density limit, so applying simple nebular theory, we  find
$n_{e}<$10$^{2-3}$cm$^{-3}$. For comparison, Walsh
\& Roy (1989) obtained $n_{e}$=10$^{2.46}$cm$^{-3}$ for the central 
$\sim 8 \times 4$ arcsec of NGC\,5253 from optical spectroscopy.

\section{Interpretation of ground and space-based IR observations}\label{sect3}

Before we undertake a detailed analysis of our ISO observations we 
first compare our ground and space-based mid-IR line fluxes. 
We also estimate the massive stellar content of NGC\,5253 from the IR dust
continuum. There are 
potential dangers in forming line ratios when the source is a galaxy, which 
can be expected to contain many regions of different  characteristics, and especially 
if it is a galaxy like NGC\,5253 which is known for 
marked spatial variations in stellar age, dust content and other features.  

\subsection{A Hot Nucleus in a Cooler Extended Core}

The infrared line fluxes listed in Table~\ref{table1} include both the 
ISO results and,
for the [S\,{\sc iv}], [Ne\,{\sc ii}], and [Ar\,{\sc iii}] lines, 
the fluxes observed from the ground
with the Irshell spectrometer (Beck et al. 1996) which had a 1.6\arcsec\ beam; 
mapping showed that the [S\,{\sc iv}] and 
[Ar\,{\sc iii}] detected by Irshell came from a region about 
2\arcsec\ in diameter.  Similar dimensions were found by Davies et al. (1998)
for the dominant Br$\gamma$ source. At the distance of NGC\,5253,
this corresponds to a (maximum) size of 40 parsec. The 
most striking thing about Table~\ref{table1} is the close agreement of the 
[S\,{\sc iv}] 10.5$\mu$m fluxes observed with the large ISO beam 
and the 1.6\arcsec\ Irshell beam.  This
result argues that the [S\,{\sc iv}] 
emission is concentrated in the inner 2--4\arcsec\ of the 
galaxy. 
However, we cannot assume that all the line fluxes are similarly concentrated; another
feature of Table~\ref{table1} is the order of magnitude excess of 
[Ne\,{\sc ii}] 12.8$\mu$m emission in the ISO observations compared to the
ground-based, which is most readily explained if the [Ne\,{\sc ii}] is extended over
an area comparable to the ISO beam so the small Irshell beam did not measure
the total. 

In Section~\ref{sect5} we shall show that [S\,{\sc iv}] emission increases
with increasing stellar temperatures and ionization parameters, with the
reverse found for [Ne\,{\sc ii}]. Therefore the most 
natural explanation is that the [S\,{\sc iv}] region, the  central
2\arcsec\ mapped by 
Irshell, contains very hot stars at high ionization parameter, while 
the [Ne\,{\sc ii}] region contains cooler stars at lower Q-value.
Other lines from high ionization stages will be likely  to 
originate in the [S\,{\sc iv}] region and those from lower ionization in 
the [Ne\,{\sc ii}] region.

This is a far-reaching assumption, but the picture of NGC\,5253 that has been
established from high spatial resolution observations supports it.  
In the radio continuum  the thermal emission due to 
young stars is 
dominated by a central source of less than 2\arcsec\ which contains more 
than half the 
radio flux from the area of the ISO beam (Turner et al. 1998). 
This radio source is extremely young, 
which has led it to be identified with the `central star cluster' 
[S\,{\sc iv}] source 
(Beck et al. 1996) and with the visually obscured cluster NGC\,5253-5 of 
Calzetti et al. (1997) 
which is thought to be no more than 2.5 Myr old, 
resides in the center of the starburst nucleus, and is very compact.  
The youth of the cluster is  seen in the lack of supernovae 
remnants and the presence of 
very massive and shortlived stars which are hot enough to excite 
strong [S\,{\sc iv}] emission. 
Calzetti et al. distinguish between the starburst nucleus, which contains 
the central cluster, and the galactic core, which is older, 
produces stars at a rate about 
1/10 that of the nucleus, is UV rich, and covers an area  comparable to 
the ISO beam.  If the [S\,{\sc iv}] source is associated with the 
starburst nucleus, the [Ne\,{\sc ii}] emission may be presumed
to come from the larger core. The halo of NGC\,5253 was not observed by 
ISO and is not discussed in this work.

\subsection{Dust Emission and the Number
of O stars in NGC\,5253}\label{dust}

The SWS spectrum of NGC\,5253  in Fig~\ref{fig1} shows that the  nebular emission 
lines are superimposed on an  underlying energy distribution  which 
has negligible flux below 4~$\mu$m, peaks at 
about 10~Jy at 24~$\mu$m and then turns over to longer wavelengths.  
This energy distribution clearly reflects emission from relatively warm
dust. 
The broad emission between 22.5--27$\mu$m suggests possible solid state 
silicate emission, a sensitive tracer of mass-loss history.
This is common in O-rich environments of the Galaxy but is not known in 
extragalactic regions, like 30~Doradus, where the dust may be insufficiently 
shielded. Therefore, NGC\,5253 appears to be the first example of 
silicate emission in a starburst galaxy.
Using blackbody energy distributions  normalised to the SWS data at about 21$\mu$m  we find 
a best  fit to the observed data for a temperature of 200$\pm$10~K, as shown in 
Fig.~\ref{fig1}. 
% (Extinctions up  to $A_{v}$=30 mag were found to have negligible effect on 
% the mid-IR continuum distribution). 
From this, the total  integrated dust emission at the earth  
is 1.6$\times$10$^{-12}$J\,s$^{-1}$\,m$^{-2}$. 
Adopting a distance  of 4.1~Mpc to NGC\,5253 (Saha et al. 1995), 
this translates into a total dust emission luminosity  of 2.9$\times$10$^{35}$J\,s$^{-1}$, or 
L$_{\rm IR}$=7.8$\times$10$^{8}$ {\lsun}. (For comparison, Lutz et al. (1996) report L$_{\rm IR}$
=8.0$\times$10$^{8}$ {\lsun} from ISO-SWS spectroscopy, while Stevens \& Strickland (1998) 
derive 6.9$\times$10$^{8}$ {\lsun} using IRAS flux measurements plus our adopted distance). 

Before we estimate the number of hot stars exciting this dust, let us
recall the concept of an ``equivalent'' O star (Vacca 1994).
This hypothetical object was introduced so as to ease the transformation
between what is measured by the Lyman continuum flux and ``the number of O
stars'' and provides a useful means of comparing the energetics of star 
formation regions in galaxies and other sources.
The number of Lyman continuum photons from an O star ``equivalent''
is $10^{49}$ Ly photon\,s$^{-1}$, close to that of an O7\,V 
star. Vacca (1994) 
calculates the correction in transforming from the number of 
``equivalent'' O stars to the ``number of O stars''. This
depends on the initial mass function (IMF) slope, the metal abundance,
and the upper mass cut-off.  Roughly half of the ionizing flux comes from
stars hotter and the other half from stars cooler than the O star
``equivalent', thus the correction factor is typically less than a factor
two. It should be kept in mind that one might be measuring the Lyman
continuum flux and determining the number of O star ``equivalents", and
estimating the ``number of O stars'' but there may only be stars of later
spectral type than O7\,V if the starburst is somewhat evolved. In fact, we
will see this case below in the ``core'' of NGC\,5253. 

\begin{table} 
\caption[]{Stellar parameters and ionizing fluxes of 
main-sequence O star flux distributions from Schaerer \& de Koter (1997),
for 0.2 $Z_{\odot}$. 
Approximate spectral types follow the temperature calibrations of 
Vacca et al. (1996, WDV) and Crowther (1997, PAC)}
\label{table2} 
\begin{center} 
\begin{tabular}{l@{\hspace{0mm}}r@{\hspace{.5mm}}r@{\hspace{2.5mm}}r
@{\hspace{.5mm}}r@{\hspace{.5mm}}r@{\hspace{.5mm}}r@{\hspace{.5mm}}r
@{\hspace{1.5mm}}l@{\hspace{1mm}}l} 
\hline 
    & Mass       & Age & $T_{\rm eff}$ & log$L_{\ast}$ 
& log$Q_{\rm H}$ & log$Q_{\rm HeI}$ & 
log$Q_{\rm HeII}$ & \multicolumn{2}{c}{Sp. Type} \\
      & $M_{\odot}$& Myr &  kK        & $L_{\odot}$    & ph\,s$^{-1}$   & ph\,s$^{-1}$   &
ph\,s$^{-1}$  & WDV & PAC \\
\hline 
A2  & 20    & 3.6  &   33.3  & 4.77 &47.84 &46.08 & 0.00 & B0\,V & O8.5\,V \\
B2  & 25    & 2.6  &   36.3  & 5.00 &48.43 &47.42 &44.23 & O9\,V & O8\,V \\
C2  & 40    & 1.5  &   41.7  & 5.45 &49.11 &48.53 &46.04 & O7\,V & O6\,V\\
D2  & 60    & 0.8  &   46.1  & 5.77 &49.51 &49.02 &46.45 & O5\,V & O4\,V\\
E2  & 85    & 0.7  &   48.5  & 6.04 &49.82 &49.36 &46.59 & O4\,V & O3\,V\\
{}F2 &120   & 0.6  &   50.3  & 6.26 &50.07 &49.62 &47.00 & O3\,V & O3\,V \\
\hline
\end{tabular}
\end{center}
\end{table}

% How to get Ly photons from observed Br_alpha?
% Br_alpha(obs)        =5.1 x 10^-16 W m^-2
% Br_alpha(E(B-V)=2.5) =6.2 x 10^-16 W m^-2  (D=1.263 x10^23 m)
% Br_alpha_intrinsic   =1.24x 10^+39 erg s^-1
% H_beta_intrinsic     =1.64x 10^+40 erg s^-1 (factor 13.26 from intrat 5-4,4-2)
% Ly photons (x2.1E12 from Vacca Conti 1992 Table 4) 
%                      =3.45x 10^+52 ph s^-1  (case B, Ne=290, Te=11 700K)

% For comparison, Davies get within 30arcsec
% Br_gamma (observed)     =9.7 E-17 W m^-2 (4arcsec aperture)  (dred) 1.5E-16 A_K=0.5
%                         =29.1E-17 W m^-2 (30 arcsec aperture) (dred) 4.5E-16 A_K=0.5
%                                                                      5.85E-16 A_K=0.7
% Br_gamma (de-reddened)  =4.60E-16 W m^-2  (A_K=0.5 => E(B-V)=1.56)
% Br_alpha (intrat x2.82) =1.30E-15 W m^-2 => 7.2 x10^52 ph s^-1
%                          1.65E-15 W m^-2 => 9.2 x10^52 ph s^-1
%

The Lyman continuum flux of a region of hot stars is usually found from the
extinction-free radio continuum, but in NGC\,5253 this is partly optically
thick and there is insufficient spatial resolution to determine which 
can be associated with the starburst nucleus and which with the core.  
% NEW 18-Aug-98
We therefore assume that the H\,{\sc ii} region is dust-free, and optically
thick to Lyman continuum radiation (Case~B ionization-bounded).
We adopt $N_{e}$=10$^{2.5}$cm$^{-3}$ and $T_{e}$=11,700K 
for the core (Walsh \& Roy 1989) for the determination of hydrogen
intensity ratios (Storey \& Hummer 1995). Our extinction-corrected 
ISO Br$\alpha$ flux implies a total ionizing flux of 
3.45$\times$10$^{52}$ Ly photon\,s$^{-1}$.
This supports the previous determination of
4$\times$10$^{52}$ Ly photon\,s$^{-1}$ from H$\alpha$ by Martin \&
Kennicutt (1995) though Davies et al. (1998) recently used 
Br$\gamma$ observations plus an aperture of 30 arcsec 
to derive a factor of two higher Lyman ionizing flux (scaled to 
our assumed distance). This discrepancy would be larger still (factor of 2.7) 
adopting a consistent IR extinction.

Davies et al. revealed a dominant nebular source,
spatially coincident with the visual H$\alpha$ emission peak, containing
30\% of the Br$\gamma$ emission within 2$''$ of the central nucleus. (A
similar fraction was measured by Martin \& Kennicutt (1995) from H$\alpha$
observations.) We therefore conclude that the ionizing flux of the nucleus,
as derived from our measured Br$\alpha$ flux, is 1.0$\times$10$^{52}$ Ly 
photon\,s$^{-1}$ -- comparable to 30~Doradus (Kennicutt
1984) -- and that the ionizing flux in the core is 
2.5$\times$10$^{52}$ Ly photon\,s$^{-1}$.
The total ionization is equal to 3\,500 equivalent O7\,V stars.

Using the recent O star calibration of Vacca, Garmany \& Shull (1996),
we find a luminosity of $\sim$2.5$\times$10$^{5}$ {\lsun} for an O7\,V star. 
With the typical assumption that the dust luminosity is governed by the radiation
from the hot stars, our derived L$_{\rm IR}$ is equivalent to about 3\,100 O7\,V
stars.  This number is in good agreement with that for the O star 
``equivalents'' inferred from the Lyman continuum measures.

\section{Photo-ionization calculations}\label{sect4}

The ISO data on NGC\,5253 include the important diagnostic lines of sulphur,
argon and neon, from whose ratios it should be possible to deduce 
the extreme-ultraviolet (EUV) spectrum of the input ionising radiation, and thus
the temperature of the exciting star (in a nebula excited by a 
single star) or the effective temperature and thus the age and mass 
function of an exciting cluster of stars.  In the following section we
will use results from a grid of model H\,{\sc ii} regions produced with the 
photo-ionization code {\sc cloudy} (Ferland 1996). 
First, we  describe our technique, choice of elemental abundances and
 O star energy distributions, since these quantities will 
have a major effect on the results.

\subsection{Description of the calculations}

We constructed photo-ionization models using {\sc cloudy} (v90.04) as
described in Ferland (1996) and Ferland et al. (1998). 
Comparisons with other photo-ionization codes are provided by
Ferland et al. (1995). The nebula are represented by a sphere
of uniform gas density, $n$, and filling factor, $\epsilon$, with
a small central cavity, which is 
ionized and 
heated solely by the UV radiation of a single central star.  Nebular fluxes
are predicted, given input abundances, flux distributions and physical 
parameters, 
most important of which is the ionization parameter. This may be written as
$Q = Q_{\rm H}/(4 \pi R^{2}_{\rm S} n)$, where $Q_{\rm H}$ is the number of
ionizing photons below
the H-Lyman edge at 912\AA, and $R_{\rm S}$ is the radius of the
Str\"{o}mgren 
sphere. 
(An alternative definition of the ionization parameter is $U = Q/c$.) 

{}For a given 
energy distribution of the ionizing radiation field, any combination of
parameters
which keeps
$Q_{\rm H} n \epsilon^2$ constant will result in an identical ionization structure of 
the gas (see Stasi\'{n}ska \& Leitherer 1996). For each flux distribution we
have followed Stasi\'{n}ska \& Schaerer (1997) by computing a series of 
models with hydrogen density $n$=10 cm$^{-3}$ and $\epsilon$=1, in which the 
stellar luminosity has been multiplied by a factor of 10$^{5}$, 10$^{4}$,
..., 10$^{-3}$. These models thus correspond to varying the ionization
parameter log~$Q$ = 6 to 10. This covers the range of ionization parameters 
expected in astophysically relevant conditions as found by Stasi\'{n}ska 
\& Leitherer (1996).

\subsection{Abundance Effects}

The elemental abundances in the ionized gas can strongly affect the cooling
of the nebula and thus the observed line ratios. NGC\,5253 is seen from
optical spectroscopy by Walsh \& Roy (1989) to be metal-poor, with 
12+log(O/H)=8.17 derived for the region centred on the major H$\alpha$ (and
Br$\gamma$) emission. More recently, Kobulnicky et al. 
(1997) have derived elemental abundances of N, O, Ne, S and Si in several regions of 
the core H\,{\sc ii} region from HST spectroscopy and confirm a depleted metal content. 
They supported earlier evidence for a region of enhanced nitrogen 
abundance, attributed to enriched material from WN stars. 
The abundance of sulphur is 1/4--1/2 times that of the Orion nebula 
(Baldwin et al. 1991). For our analysis we use 
abundances derived for region H\,{\sc ii}--1 by Kobulnicky et al. (1997), and 1/3 
that of the Orion nebula  (or approximately 1/4 solar metallicity) for 
elements Kobulnicky et al. were unable to measure.

Test calculations were performed including varying 
of dust grains abundances and compositions (ISM, Orion).
Overall the influence of dust was not substantial, with the 
greatest differences obtained in high luminosity models, such that 
the ionization balance and nebular temperature increase. 
We found that the optimum consistency between different
IR diagnostics was obtained for models  including ISM grains 
(silicate and graphite). Consequently, all models considered in 
this work assume this grain mixture.

% O/H =1.47E-4 WR region 1       1.31E-4 Kobulnicky HII-1
% N/H = 1.13E-5 WR region 1      1.9E-5  
% Ne/H=1.98E-05 WR region 1      2.38E-5
% S/H=4.35E-06 WR region 1       3.635E-6 Kon HII-1  6.53E-6 in HII-2
% Si/H=                          1.94E-6 

% S/H=1.33E5 in Orion from Baldwin et al. 1991 (S/H=1.88E-5 SOLAR) => NGC\,5253 = 0.19-0.34

\subsection{Flux distributions for O-type stars}

Most of the recent studies on analysis of H\,{\sc ii} regions use the Kurucz (1991)
plane-parallel LTE model atmospheres. This is because these models include line
blanketing and are available for a wide range of stellar temperatures and metallicities.
However, with regard to O-type stars, Kurucz models neglect both non-LTE effects and
the influence of stellar winds, with important consequences for the EUV
flux distribution from such stars (e.g. Sellmaier et al. 1996; Schaerer \& 
de Koter 1997). 

We have therefore used the grid of theoretical stellar flux distributions 
from Schaerer \& de Koter (1997). In contrast with the
Kurucz (1991) O star flux distributions, those of Schaerer \& de Koter (1997) 
(i) differ greatly below the He\,{\sc ii} $\lambda$228 edge, and (ii)
show little metallicity dependence in the important $\lambda\lambda$228--912 region.
Indeed, Kurucz (1991) and Schaerer \& de Koter (1997) flux distributions
show only minor differences at low metallicities in this region. 
Stasi\'{n}ska \& Schaerer (1997) have demonstrated the improved agreement 
between theoretical predictions and observations of H\,{\sc ii} regions for
Schaerer \& de Koter (1997) fluxes compared to those from Kurucz (1991).

The Schaerer \& de Koter (1997) have an 
additional advantage over Kurucz in that they are coupled to the Geneva
stellar evolution code, so that each flux distribution relates to a
particular age, luminosity and mass. We shall therefore use the
main-sequence (A2--F2) models  from Schaerer \& de Koter (1997),
parameters of which are listed in Table~\ref{table2}. 
These were calculated with a metallicity 
of 1/5 Z$_{\odot}$, appropriate to the metal content of NGC\,5253.
Two temperature entries provide equivalent O spectral types from these 
models: Vacca et al. (1996) constructed  an empirical $T_{\rm eff}$--calibration 
based on observational and theoretical results available at the time. An updated OB
calibration was provided by Crowther (1997) including more recent 
results, and omitting the linear $T_{\rm eff}$--spectral type relationship 
that was assumed by Vacca et al. (1996).

\subsection{Extinction Effects}\label{extinction}

We cannot assign a single extinction to the central region of NGC\,5253, because the
extinction towards the center is clearly very patchy and irregular (as can be 
seen in the dust lane and in the $H\alpha/H\beta$ ratio map of Calzetti et al 1997).  
Observers have found extinction values which range from $A_v\le1$ to
$A_v=35$~mag
depending on the wavelength, aperture, and exact location of the observation. 
The center of the galaxy is an association of many small regions of greatly 
differing extinctions; we have here to determine reasonable values of 
extinction for the hot nucleus and the cooler core.  

Calzetti et al. (1997) find that the true nucleus, which we have identified with the
[S\,{\sc iv}] source, is quite opaque in the visible; comparing the $H\alpha$ flux to
the 2-cm radio they found that the nuclear star cluster is embedded in a cloud
of total $A_v$=9--35 magnitudes.   The Brackett $\alpha$ and $\gamma$ lines
of Kawara et al (1989) give $A_v$ of 7--12 magnitudes in an area of 10$\times$20 arcsec. 
The extinction to the rest of the core region is
much lower, such that the total $H\alpha$ compared to the 6-cm flux in the inner
38\arcsec\ gives $A_v$ of about 1. This agrees with the model that the youngest
stars in NGC\,5253 are in the nucleus; it is expected that the youngest stars will be 
the more highly obscured as they have not had time to disperse the dense molecular 
cloud in which they formed.

{}For this work we shall use the Kawara et al. (1989) Br$\gamma$/Br$\alpha$
ratio since the region sampled by Kawara et al. is comparable to the ISO
aperture. The relation  between the extinction in the infrared and $A_v$ is 
also not well established. Nevertheless, one magnitude of $9.7\mu$m extinction is 
estimated to be between 15 and 30 magnitudes $A_v$. Although the extinction in 
the infrared is much lower than in the optical it can still have significant 
effects on the line ratios.  Extinction curves in the infrared differ from 
source to source, but in broad terms the [S\,{\sc iv}] and [Ar\,{\sc iii}] lines will 
be most affected, and similarly affected, by extinction. From the Kawara et al.
Br$\gamma$/Br$\alpha$ ratio we find $A_{v}$=7.7 mag,
$A_{\rm K}$=0.7 mag and $A_{\rm M}$=0.2 mag using the Howarth 
(1983) extinction curve. Consequently, the extinction at $9.7\mu$m, is
0.25--0.5 mag, and from the curve of Draine (1989)
there is 0.2--0.4 mag extinction at [Ar\,{\sc iii}] and [S\,{\sc iv}],
0.1--0.2 mag
at [S\,{\sc iii}], 0.08--0.17 mag at [Ne\,{\sc ii}], and 0.06--0.13 mag at 
[Ne\,{\sc iii}].  
% Since, as we will show below, the [Ar\,{\sc iii}] and
% [S\,{\sc iv}] lines are likely to come from the same region and thus
% undergo the same reddening, the problem is much simplified. 

\section{Results of photo-ionization modelling}\label{sect5}

\begin{figure*}
\vspace{14.4cm}
\includegraphics{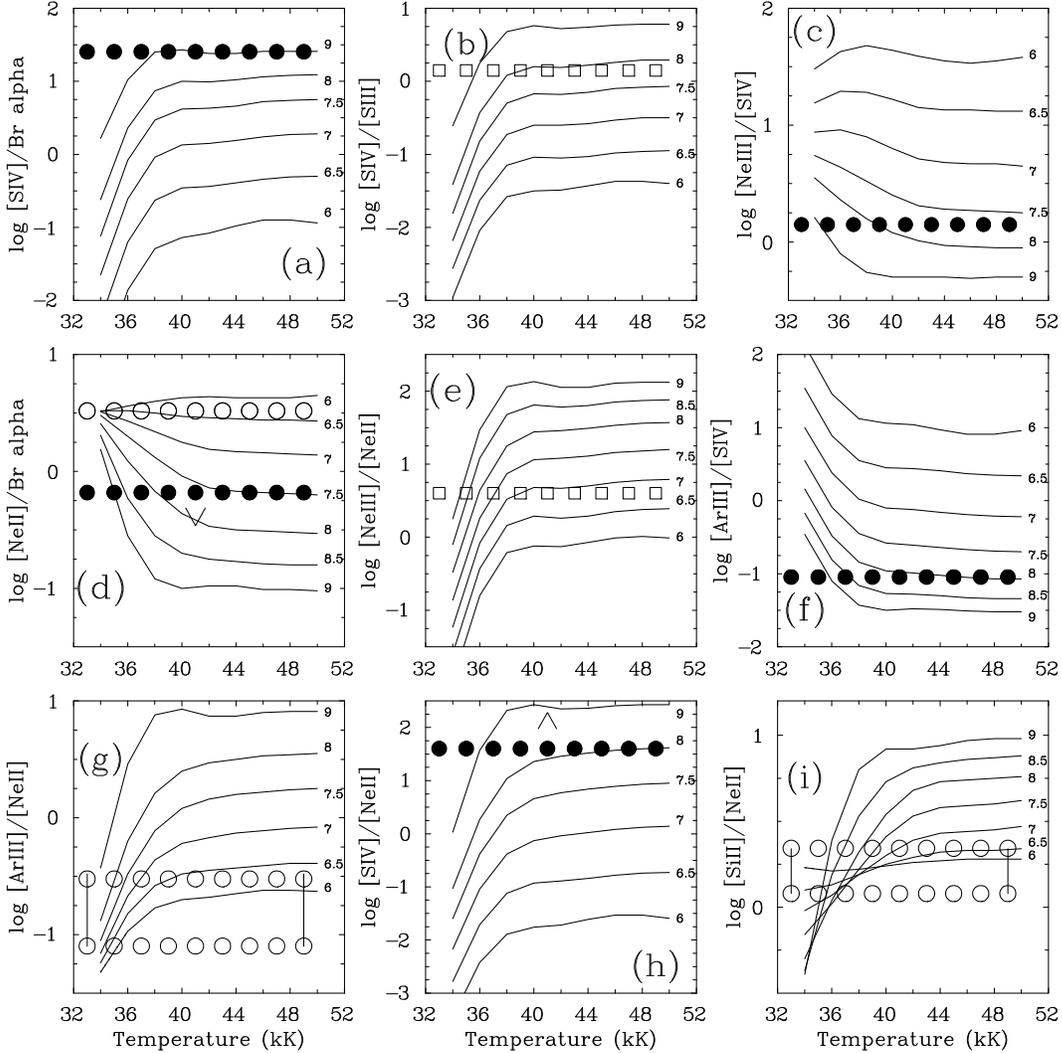}
\caption{Sensitivity of selected IR line ratios to ionization
parameter ($Q$) and stellar temperature ($T_{\rm eff}$).
Circles indicate observed line ratios in the nucleus (filled-in)
and core (open), while squares indicate ratios for the 
combined core--nucleus region. Some observed ratios lie within 
a particular range of values (connected by solid lines) while
others are merely limits (indicated by arrows).}
\label{fig3}
\end{figure*}

Following the approach described above, we now proceed to 
determine the massive stellar content in both the starburst 
nucleus and core of NGC\,5253.
Fig~\ref{fig3} shows how selected IR line ratios are affected by
changes in temperature and ionization parameter. For a 
given density,  the dependance of line
ratios on $Q$ translates into a sensitive function of the input 
ionization, and thus on the stellar temperature. It is apparent that 
[S\,{\sc iv}] emission is predicted at high  
$Q$ and $T_{\rm eff}$, while [Ne\,{\sc ii}] originates from lower
$Q$ and $T_{\rm eff}$ regions. 
(Almost identical contours are obtained using the {\it solar} metallicity 
theoretical flux distributions of Schaerer \& de Koter (1997)).
We shall now exploit these dependances to derive characteristic 
stellar temperatures and nebular densities for the two regions under 
consideration.

% Results for predicted IR line strengths for model~1, which has $\log~Q$=9--9.5, 
% are shown in Table~\ref{table3}. This is the same ionization structure as 
% a model with a single ionizing star and gas density 10$^{6}$ cm$^{-3}$. To 
% demonstrate the 
% effect of lower ionization parameters, we include 
% models~3 and 5, with ionization structures equivalent 
% to that 
% of a single star with gas density of 10$^{2}$ and 10$^{-2}$ 
% cm$^{-3}$, respectively, in Table~\ref{table3}, 
% Line fluxes are presented relative to Br$\alpha$=1.0. (In these units, 
% H$\beta$=11.4--12.8, over the range of densities and temperatures under 
% consideration (Storey \& Hummer 1995). 

\begin{table} 
\caption[]{{\bf (a)} Results from photo-ionization models of individual O stars
(Schaerer \& de Koter 1997) that are 
consistent with selected diagnostic IR line ratios for the 
\underline{nucleus} of NGC\,5253. Preferred solutions are indicated by $\bullet$}
\label{table3a} 
\begin{center} 
\begin{tabular}{
l@{\hspace{3mm}}
l@{\hspace{3mm}}
r@{\hspace{3mm}}
r@{\hspace{3mm}}
r@{\hspace{3mm}}
r@{\hspace{3mm}}
r@{\hspace{3mm}}
r@{\hspace{3mm}}
r} 
\hline 
$T_{\rm eff}$  & log~$Q$  & \underline{Ar\,{\sc iii}} & \underline{S\,{\sc iv}}& \underline{Ne\,{\sc iii}}&\underline{S\,{\sc iv}} &\underline{Ne\,{\sc iii}}&\underline{S\,{\sc iv}} &\\
kK             &          & S\,{\sc iv}               & Br$\alpha$             & S\,{\sc iv}              &Ne\,{\sc ii}              &Ne\,{\sc ii}             & S\,{\sc iii} &\\
\hline 
\multicolumn{2}{c}{Observed}& 0.09                    & $\sim$25               &$\le$1.4                       & $>$39                & $>$46?                & $>$1.4 &\\
\hline
%36             & 8.75     &   0.10                    &  8                     & 1.0                      &19                      & 19                      & 1.3 &\\
36             & 9.0      &   0.08                    & 11                     & 0.8                      & 37                     & 29                      & 1.8 &\\
\noalign{\smallskip}
%38             & 8.0      &   0.09                    &  7                     & 1.6                      & 11                     & 18                      & 1.2 &\\
38             & 8.25     &   0.10                    & 11                     & 1.1                      &  26                    & 30                      & 1.8 &\\
38             & 8.5      &   0.07                    & 15                     & 0.9                      &  55                    & 48                      & 2.6 &$\bullet$ \\
38             & 8.75     &   0.05                    & 20                     & 0.7                      & 115                    & 76                      & 3.5 &\\
\noalign{\smallskip}
42             & 7.75     &   0.15                    &  7                     & 1.4                      &  13                    & 19                      & 1.0 &\\
%42             & 8.0      &   0.10                    & 10                     & 1.0                      &  29                    & 29                      & 1.5 &\\
42             & 8.25     &   0.09                    & 15                     & 0.7                      &   56                   & 48                      & 2.5 &$\bullet$\\
%42             & 8.5      &   0.05                    & 17                     & 0.6                      &  98                    & 60                      & 3.2 & \\
42             & 8.75     &   0.04                    & 21                     & 0.5                      &  155                   & 83                      & 4.2 &\\
\noalign{\smallskip}
46             & 7.75     &   0.13                    &  8                     & 1.3                      &   18                   & 22                      & 1.2 &\\
%46             & 8.0      &   0.09                    & 11                     & 0.9                      &  37                    & 34                      & 1.8 &\\
46             & 8.25     &   0.06                    & 15                     & 0.7                      &   71                   & 50                      & 2.6 &$\bullet$\\
%46             & 8.5      &   0.05                    & 19                     & 0.6                      & 120                    & 71                      & 3.6 &\\
46             & 8.75     &   0.04                    & 23                     & 0.5                      &  182                   & 98                      & 4.7 &\\
\noalign{\smallskip}
50             & 7.75     &   0.13                    &  9                     & 1.2                      &   20                   & 24                      & 1.3 &\\
%50             & 8.0      &   0.09                    & 12                     & 0.9                      &  42                    & 37                      & 1.9 &\\
50             & 8.25     &   0.06                    & 16                     & 0.7                      &   78                   & 54                      & 2.8 &$\bullet$\\
%50             & 8.5      &   0.05                    & 20                     & 0.6                      & 129                    & 76                      & 3.8 &\\
50             & 8.75     &   0.04                    & 23                     & 0.5                      &  190                   &100                      & 4.8 &\\
\hline
\end{tabular}
\end{center}
\end{table}

\addtocounter{table}{-1}

\begin{table} 
\caption[]{{\bf (b)} Identical to Table~3a, except for the
\underline{core} of NGC\,5253.}
\label{table3b} 
\begin{center} 
\begin{tabular}{
l@{\hspace{3mm}}
l@{\hspace{3mm}}
c@{\hspace{3mm}}
r@{\hspace{3mm}}
c@{\hspace{3mm}}
r@{\hspace{3mm}}
r@{\hspace{3mm}}
r} 
\hline 
$T_{\rm eff}$  & log~$Q$  & \underline{Ar\,{\sc iii}} & \underline{Ne\,{\sc ii}}& \underline{Si\,{\sc ii}}&\underline{S\,{\sc iii}}&\underline{S\,{\sc iv}}&\\
kK             &          & Ne\,{\sc ii}               & Br$\alpha$             & Ne\,{\sc ii}              &Ne\,{\sc ii}             & Ne\,{\sc ii} &\\
\hline 
\multicolumn{2}{c}{Observed}& 0.08--0.3               & $\sim$3.3             & 1.2--2.2                & $\le$2                  & $<$0.35 &\\
\hline
34             & 7.25     &   0.08                    & 3.2                     & 0.8                      & 1.3                      & 0.01                      &\\
34             & 8.0      &   0.13                    & 2.6                     & 0.5                      & 2.3                      & 0.09                      &\\
34             & 8.75     &   0.29                    & 1.8                     & 0.4                      & 3.9                      & 0.62                      &\\
\noalign{\smallskip}
35             & 7.0      &   0.13                    & 3.0                     & 1.0                      & 1.3                      & 0.03                      &\\
35             & 7.5      &   0.17                    & 2.5                     & 0.9                      & 2.1                      & 0.11                      &\\
35             & 8.0      &   0.31                    & 1.7                     & 0.8                      & 3.5                      & 0.47                      &\\
\noalign{\smallskip}
36             & 6.0      &   0.11                    & 3.6                     & 1.6                      & 0.4                      & 0.00                      &\\
36             & 6.5      &   0.15                    & 3.3                     & 1.3                      & 0.8                      & 0.02                      &\\
36             & 7.0      &   0.21                    & 2.7                     & 1.2                      & 1.5                      & 0.09                      &\\
\noalign{\smallskip}
37             & 6.25     &   0.17                    & 3.5                     & 1.5                      & 0.7                      & 0.02                      &\\
\hline
\end{tabular}
\end{center}
\end{table}

\subsection{[S\,{\sc iv}] emitting central nucleus}

In  Section~\ref{sect3} we argued that the [S\,{\sc iv}] emission is 
concentrated in the nucleus region of NGC\,5253 observed by Beck et 
al. (1996). Our principal nucleus diagnostics are therefore: 
(i) [Ar\,{\sc iii}]/[S\,{\sc iv}]
from Beck et al. (1996); (ii)
[S\,{\sc iv}]/[Ne\,{\sc ii}] from Beck et al. (1996);
(iii) [S\,{\sc iv}]/Br$\alpha$, given that $\sim$30\% of the 
total ISO Br$\alpha$ flux is emitted in the nucleus (Section~\ref{dust}); 
(iv) an upper limit on [Ne\,{\sc iii}]/[S\,{\sc iv}] from ISO, since 
[S\,{\sc iv}] is formed exclusively in the core; (v) 
a lower limit on [S\,{\sc iv}]/[S\,{\sc iii}] 
for similar reasons. [Ne\,{\sc iii}]/[Ne\,{\sc ii}] represents a 
secondary diagnostic since it relies on assumptions regarding the formation 
of [Ne\,{\sc iii}]  in the nucleus.

In Fig.~\ref{fig3}, filled-in circles indicate the position of 
these IR line ratios in the nucleus. Each diagnostic supports
a high ionization parameter for the nucleus. For example, 
[Ar\,{\sc iii}]/[S\,{\sc iv}] requires solutions in the
range ($T_{\rm eff}$/kK, $\log~Q$)=(36, 9) to (50, 8) from 
Fig.~\ref{fig3}(f). The observed [S\,{\sc iv}]/Br$\alpha$ ratio 
suggests even higher ionization parameters from  Fig.~\ref{fig3}(a).

Unfortunately, despite our best efforts, 
no single solution exists for which all observed line ratios are satisfied.
As previous investigators have also found, it is extremely difficult
to break the degeneracy between ionization parameter and hardness of 
the ionizing spectrum from modelling alone.
At most two or three are in agreement, suggesting errors in our assumptions
relating to line formation regions, photo-ionization modelling or flux
distributions. Table~\ref{table3a} compares selected IR diagnostic ratios
at particular temperatures and ionization parameters with observed ratios
for the nucleus. We find that optimum agreement is obtained for a narrow
`corridor' ranging in ($T_{\rm eff}$/kK, $\log~Q$) from (38, 8.5) to 
(50, 8.25). Each of these solutions reproduce the observed nebular 
properties equally well. 
Consequently, the nucleus is dense and almost certaintly contains very early
O-type stars.
The minimum nucleus stellar temperature corresponds to an O7\,V star
and higher temperatures to earlier types. As stated above, the ionization 
requires 1\,000 equivalent O7\,V stars; if the actual stars are younger,
fewer will be needed. 

Our principal diagnostic ratios have the basic problem that they combine 
ions of two different elements and may therefore be 
in error if the elemental abundances determined by Kobulnicky et al. (1997) 
are incorrect. This is why ratios of lines from different ions of the 
same element are usually to be preferred. However, in NGC\,5253 the two 
emission regions appear to be in such extreme states of temperature and 
ionization parameter that the contribution of one ionic line completely 
dominates.

% Weak [O\,{\sc iv}] emission may be present in NGC\,5253. It is reasonable to
% assume that this would be restricted to the high temperature nucleus.
% For our derived ionization parameter, the maximum 
% [O\,{\sc iv}]/[S\,{\sc iv}] ratio predicted is $\sim$0.005 for our
% 50kK model, while a ratio of $\le$0.02  is observed.  
% In Section~\ref{sect6} we evaluate the possible contribution of 
% WR stars to this feature. 

\subsection{[Ne\,{\sc ii}] emitting core}\label{Ne2_core}

We now discuss constraints on the stellar content and density of 
the [Ne\,{\sc ii}] emitting core region of NGC\,5253. Since
the [Ar\,{\sc ii}] line flux is uncertain (Sect.~\ref{IRfluxes})
we are unable to use the [Ar\,{\sc ii}]/[Ne\,{\sc ii}] ratio ($\le$0.35) 
to constrain the core properties. All solutions lie in 
the range 0.15$\le$[Ar\,{\sc ii}]/[Ne\,{\sc ii}]$\le$0.36, assuming a 
Ne abundance from Kobulnicky et al.  (1997) and Ar abundance of 1/4 solar, 

Consequently, our choice of diagnostics for the core region are: 
(i) [Ne\,{\sc ii}]/Br$\alpha$ 
from ISO, adjusted for the presence of Br$\alpha$ in the nucleus;  (ii)
[Ar\,{\sc iii}]/[Ne\,{\sc ii}], adjusted for the presence of 
[Ar\,{\sc iii}] in the nucleus and the maximum range in line fluxes; (iii)
[Si\,{\sc ii}] 34.8$\mu$m/[Ne\,{\sc ii}], using ISO
measurements from Genzel et al. (1998), assuming the Si abundance from 
Kobulnicky et al. (1997). Note that these fluxes were obtained from 
different sized apertures, so this ratio should be 
treated with caution; (iv) An upper limit on [S\,{\sc iii}]/[Ne\,{\sc ii}].
(In reality, [S\,{\sc iii}] is anticipated 
to form in both the nucleus and core regions, analogous to 
[Ar\,{\sc iii}]). We must wait for mid-IR instruments such as 
{\sc michelle} for observational verification of these predictions.
The [S\,{\sc iv}] flux originating in the core is formally
$<1 \times$10$^{-16}$W\,m$^{-2}$, obtained by subtracting the nucleus 
[S\,{\sc iv}] flux of Beck et al. (1996) from the  maximum possible
ISO flux. However considering sources of uncertainty, a strict upper 
limit of $<4\times$10$^{16}$W\,m$^{-2}$ is chosen, so that we may
additionally use an upper limit on the [S\,{\sc iv}]/[Ne\,{\sc ii}] ratio.

Observed IR line 
ratios for the core are indicated by open circles in Fig.~\ref{fig3}
while Table~\ref{table3b}(b) summarises selected ($T_{\rm eff}, Q$)
combinations. Once again, no single solution exists for which 
all our diagnostics are reproduced. Nevertheless, a very low 
stellar temperature of 34--36kK is favoured for the ionizing O stars, 
with the ionization parameter poorly constrained. 

In order to produce sufficient Lyman continuum 
ionizing photons for the observed Br$\alpha$ flux in the core
requires 2\,500 equivalent O7\,V stars (Sect.~\ref{dust})
-- significantly more if the stars are much cooler, as appears 
to be the case: $T_{\rm eff}\sim$35kK corresponds to a stellar 
type of O8 or O9 (Table~\ref{table2}). Table~\ref{table3b}(b) 
appears to permit ionization parameters as low as log~Q$\sim$6. However,
for this Q-value, the entire volume of the core would only just 
be sufficient to
contain the H\,{\sc ii} regions associated with 2,500 O7\,V equivalents.
Therefore a higher ionization parameter is favoured, albeit probably lower 
than the nucleus of NGC\,5253. Stasi\'{n}ska  \& Leitherer (1996) obtained a 
fairly tight spread of ionization parameter for young starburst regions.

% \subsection{Shock photo-ionization modelling}

Combining the above material gives a picture of a hot, dense nucleus with 
$T_{\rm eff} \ge 38$kK and $\log\,Q \sim 8.25$, containing 
1\,000 O7\,V ``equivalents'', plus a cooler core of $T_{\rm eff}$$\sim$35kK
and $\log~Q \le$8, with 2\,500 O7\,V ``equivalents''. It should be 
noted that the core is cool only by comparison with the nucleus; in a 
normal star forming system the low ionization lines of [Ne\,{\sc ii}] and 
[Ar\,{\sc ii}] dominate the mid-IR spectrum, which is not the case 
in NGC\,5253. 

Our results thus far have relied on the validity of the 
photo-ionization model {\sc cloudy}. Although this code is 
widely employed, and has been verified against a suite 
of other photo-ionization codes (Ferland et al. 
1995), we have performed additional calculations using 
the shock photo-ionization code {\sc mappings} 
(Sutherland \& Dopita 1993). The {\sc mappings} results agree 
with the overall picture of the hot nucleus in the cooler core, 
but find rather higher ionization parameters and stellar temperatures 
for both regions. We believe this discrepancy can be principally 
attributed to the use of different O star flux distributions 
(Kurucz in the case of {\sc mappings}) and may serve as a
reminder -- and measure -- of the disagreements currently 
inherent in the use of different photo-ionization models.
To verify our results further we have performed additional calculations,
using theoretical Wolf-Rayet flux distributions.

\begin{figure}
\vspace*{17.0cm}
\includegraphics{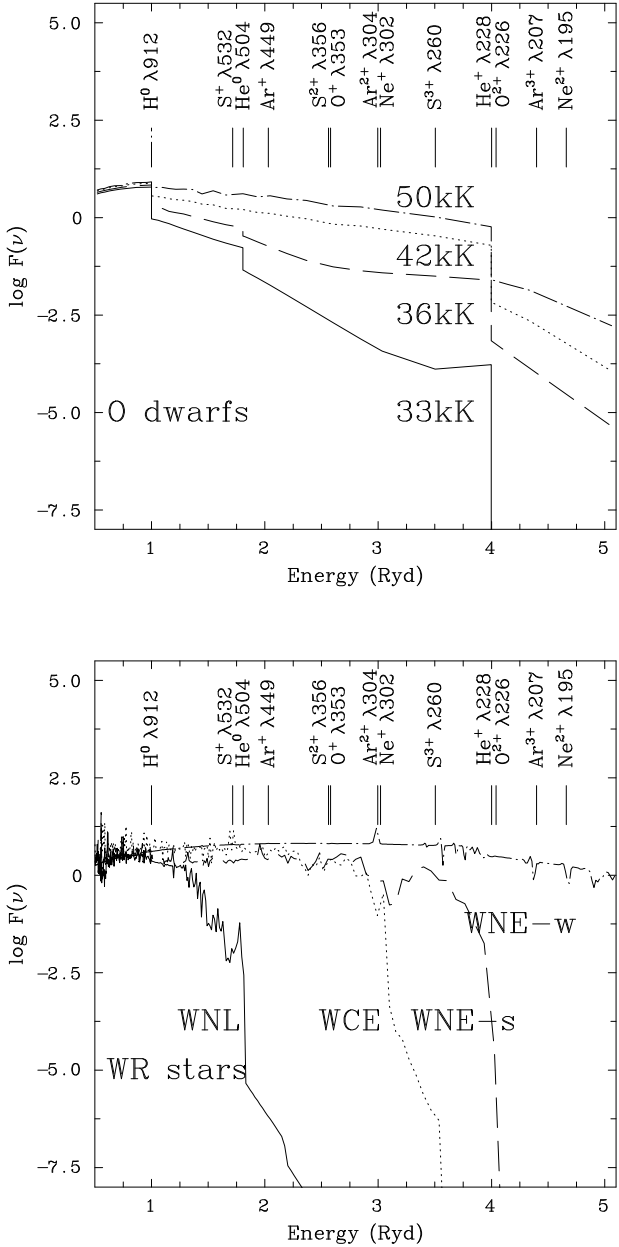}
\caption{Comparison between latest Lyman continuum flux distributions from O stars
(Schaerer \& de Koter 1997) at low metallicity (0.2$Z_{\odot}$) with recent line blanketed model
atmospheres for Wolf-Rayet stars (following Hillier \& Miller 1998). It is apparent
that strong-lined WCE and WNE stars show similar EUV flux distributions to hot O stars, except 
that negligible flux is emitted at high energies (3--4 Ryd), while solely the WNE-w model
shows a strong flux above 4 Ryd ($\lambda \le 228$\AA).
Ionization edges relevant to this study are indicated.}
\label{fig4}
\end{figure}

\section{Contribution of Wolf-Rayet stars}\label{sect6}

Up to now we have assumed that O stars are solely responsible for the 
Lyman ionizing flux from the entire galactic core. 
However, NGC\,5253 is a Wolf-Rayet galaxy, with both WN-type and WC-type stars 
present (Schaerer et al. 1997). Schaerer et al. estimated 
a population of $\le$25 WN stars and 10 WC stars in the nucleus sampled by
the ground-based IR data sets of Beck et al. (1996), with an additional 40 WN
stars and 13 WC stars located 3$''$ (60pc) to the south (also within the ISO beam). 
Schmutz, Leitherer \& Gruenwald (1992) have
indicated that Wolf-Rayet stars produce a harder flux
distribution than O-type stars, with significant energy emitted 
shortward of $\lambda$=228\AA, except for those stars 
with the strongest winds. Is it possible that the WR stars of
NGC\,5253, though few in number, contribute significantly to the high
excitation nebular lines (e.g. [S\,{\sc iv}]), and invalidate our earlier
conclusions?

\subsection{Wolf-Rayet flux distributions -- what role for line blanketing?}

The widely used WR energy distributions of Schmutz et al.
(1992) employ non line-blanketed, pure helium, model atmospheres. 
Models including light metals (e.g. CNO elements) are also in
use (e.g. Crowther, Smith  \& Hillier 1995). Line blanketing  by heavy
elements (especially Fe) is known to dramatically affect the emergent
extreme UV
flux distributions (see e.g. Crowther, Bohannan \& Pasquali 1998),
but only test model calculations with heavy metals have been 
published to date (Schmutz 1997; Hillier \& Miller 1998), 
and large grids not yet available.  We have therefore calculated
representative line-blanketed models
for late WN-type (WNL), strong lined early WN-type (WNE-s) and early 
WC-type (WCE) stars at low metallicities  ($\sim$0.2 $Z_{\odot}$)
using the Hillier \& Miller (1998) approach. We have omitted calculations
for late WC stars since they are not considered to exist in 
low metallicity environments. The reader is referred to Hillier \& Miller
(1998) for details of the technique used, except that our calculations 
include hydrogen, helium, carbon, nitrogen (for WN stars), oxygen 
(for WC stars) and iron.

% 13-Jul-98 - WNE-w line blanketed model included
\begin{table} 
\caption[]{Comparison between IR line ratios predicted in photo-ionization
modelling of NGC\,5253 using line blanketed O-type models 
(Schaerer \& de Koter 1997) and WR distributions (following Hillier  \& 
Miller 1998) at 0.2$Z_{\odot}$. The O 
star $T_{\rm eff}$--spectral type calibration of Crowther (1997) is adopted.}
\label{table4} 
\begin{center} 
\begin{tabular}{l@{\hspace{0.5mm}}r@{\hspace{1.5mm}}r
@{\hspace{1.5mm}}r@{\hspace{1.5mm}}r@{\hspace{1mm}}r
@{\hspace{0.5mm}}r@{\hspace{0.5mm}}r@{\hspace{0.5mm}}r} 
\hline
& \multicolumn{4}{c}{O dwarfs} & \multicolumn{4}{c}{Wolf-Rayet stars} \\
Sp Type &  O8.5 &  O7.5 & O6 &  O3 & WNL  & WNE-s  & WNE-w & WCE \\
%           33.5       37.5       42        50
\hline
 & \multicolumn{8}{c}{ $\log~Q$=8.5 } \\
Ne\,{\sc iii}/Ne\,{\sc ii}&0.4 &38.0 &60.3 &75.9 & 0.0  & 69.1 & 99.9 & 1.7 \\
S\,{\sc iv}/S\,{\sc iii}  &0.1 & 2.2 & 3.2 & 3.8 & 0.0  &  4.3 &  4.9 & 1.7 \\
Ar\,{\sc iii}/Ar\,{\sc ii}&0.5 &21.9 &24.0 &23.4 & 0.0  & 28.2 & 15.2 &50.2 \\
Ne\,{\sc ii}/Br$\alpha$   &2.9 & 0.3 & 0.2 & 0.2 & 3.3  &  0.2 &  0.1  & 3.0 \\
S\,{\sc iv}/Br$\alpha$    &0.4 &13.2 &17.4 &20.0 & 0.0  & 23.5 & 23.0 &13.0 \\
O\,{\sc iv}/Br$\alpha$    &0.0 & 0.2 & 0.2 & 0.5 & 0.0  &  0.0 & 49.4  & 0.0 \\
 & \multicolumn{8}{c}{ $\log~Q$=7.5 } \\
Ne\,{\sc iii}/Ne\,{\sc ii}&0.1 & 5.9 &12.0 &15.8 & 0.0  & 15.1 & 29.8 & 0.3 \\
S\,{\sc iv}/S\,{\sc iii}  &0.0 & 0.4 & 0.7 & 0.8 & 0.0  &  1.0 &  1.7 & 0.3 \\
Ar\,{\sc iii}/Ar\,{\sc ii}&0.2 & 3.7 & 9.3 &10.2 & 0.0  & 12.6 & 8.9 &13.6 \\
Ne\,{\sc ii}/Br$\alpha$   &3.3 & 1.5 & 0.7 & 0.6 & 2.7  &  0.7 & 0.4 & 4.1 \\
S\,{\sc iv}/Br$\alpha$    &0.0 & 2.4 & 4.3 & 5.6 & 0.0  &  7.0 &10.8 & 2.5 \\
O\,{\sc iv}/Br$\alpha$    &0.0 & 0.0 & 0.0 & 0.1 & 0.0  &  0.0 & 21.6  & 0.0 \\
 & \multicolumn{8}{c}{ $\log~Q$=6.5 } \\
Ne\,{\sc iii}/Ne\,{\sc ii}&0.0 & 1.2 & 1.8 & 2.5 & 0.0  & 2.4  & 5.5 & 0.1 \\ 
S\,{\sc iv}/S\,{\sc iii}  &0.0 & 0.1 & 0.1 & 0.1 & 0.0  & 0.1  & 0.3 & 0.0 \\
Ar\,{\sc iii}/Ar\,{\sc ii}&0.1 & 1.2 & 2.3 & 2.6 & 0.0  & 3.3  & 2.8  & 2.9 \\
Ne\,{\sc ii}/Br$\alpha$   &3.3 & 3.2 & 2.9 & 2.7 & 1.4  & 3.0  & 2.1  & 5.0 \\
S\,{\sc iv}/Br$\alpha$    &0.0 & 0.2 & 0.4 & 0.5 & 0.0  & 0.7  & 1.5  & 0.2 \\
O\,{\sc iv}/Br$\alpha$    &0.0 & 0.0 & 0.0 & 0.0 & 0.0  & 0.0  & 4.0  & 0.0 \\
\hline
\end{tabular}
\end{center}
\end{table}

{}Fig.~\ref{fig4} presents a comparison between the EUV
flux distributions of  these WR models with the Schaerer
\& de Koter (1997) models of low metallicity dwarf O stars. WNE-s  models 
incorporating
line blanketing show flux  distributions in the 228--912\AA\ range
which  are fairly similar to the 50kK O star model of  Schaerer
\& de Koter (1997) at low metallicity. The strong, line-blanketed
WR stellar wind hinders emission below $\lambda$=228\AA, in 
contrast with the most recent recent O star models.  
WCE models are similar except that negligible flux is emitted $\lambda 
\le$300\AA, due to the blanketing by carbon and oxygen. 
WNL stars show a much softer energy distribution.

Our results appear in conflict with the much harder fluxes predicted
by certain unblanketed models of Schmutz et al. (1992), 
particularly below the He$^{+}$ edge. However, 
Schmutz et al. correctly stress the importance of stellar wind 
density, such that emission at $\lambda \le$228\AA\ relies on the WR 
wind being relatively transparent. Denser winds destroy photons
beyond this edge. Does this remain valid for metal-line blanketed 
models?  In Fig.~\ref{fig4} we include the predicted
EUV spectrum for a line blanketed model of the extremely
weak-lined, early-type WN (WNE-w) star HD\,104994 (WR46, WN3) using 
parameters from Crowther et al. (1995), except for 
a low Fe-abundance of  0.2$Z_{\odot}$. In agreement with Schmutz et al.
the flux distribution for this star is in dramatic contrast with other 
WR (and O star) line blanketed models in that 
it displays  a very hard spectrum above the He$^{+}$ edge.
This is be expected, since a subset of WR stars do emit a significant 
flux shortward at $\lambda \le$228\AA, as evidenced from nebular 
He\,{\sc ii} $\lambda$4686  emission (e.g. Dopita 
et al. 1990; Garnett et al. 1991).

To test for the influence of WR stars we have carried out {\sc cloudy}
calculations similar to those discussed above. In Table~\ref{table4} 
predictions for selected IR line ratios at $\log~Q$=6.5, 7.5 and 8.5 
using our WR energy distributions are contrasted with those from 
a variety of O star models. These results are now discussed below.

For all realistic ionization parameters, WNL (specifically WN8) stars 
have negligible contribution to the high excitation nebular lines 
observed in  the nucleus of NGC\,5253, as may be expected from Fig.~\ref{fig4}.
WNL stars do, however, make contributions to
the low excitation features ([Ne\,{\sc ii}], [Si\,{\sc ii}]) in
the core of NGC\,5253 that are comparable to late O dwarfs. 
We can safely conclude that the presence of 40--65 WNL stars 
in NGC\,5253 (according to Schaerer et al. 1997) will not 
significantly affect our earlier results based solely on O stars.

The similarity in the EUV flux distributions of WNE-s and WCE models to
early O dwarfs is reflected in their comparable diagnostic line ratios
of neon and sulphur. Consequently, the presence of $\sim$23 WC stars 
in NGC\,5253 will not influence our results for either region. 

In general, the presence of WR stars in young starburst regions do not 
affect results from exclusively O stars (however see below). WNL flux 
distributions do not differ significantly from very late O stars, 
while WNE-s and WCE energy distributions closely mimic early-type O stars. 
As we shall see in Sect.~\ref{sect7}, this is in sharp contrast with 
recent evolutionary synthesis models (Schaerer \& Vacca 1998) which utilise 
the WR models of Schmutz et al. (1992) 
showing strong emission at $\lambda \le$228\AA.

\begin{figure}
\vspace*{14.0cm}
\includegraphics{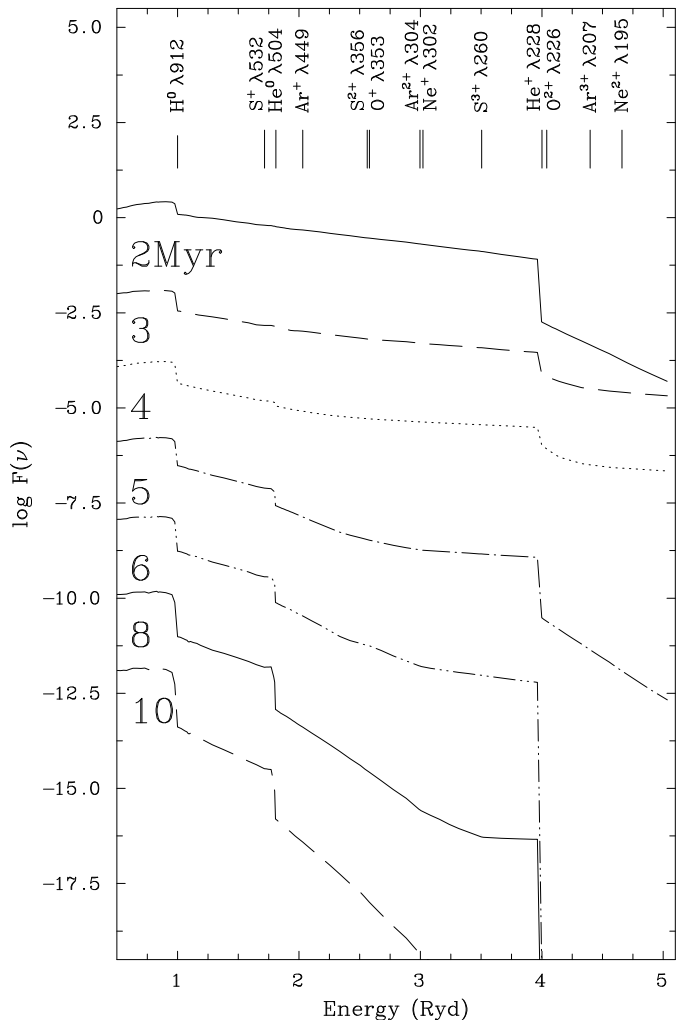}
\caption{Predictions of flux distributions for a burst of
star formation at an age of 2--10\,Myr from Schaerer \& Vacca (1998) at Z=0.2$Z_{\odot}$, 
using the Meynet et al. (1994) evolutionary tracks coupled with O star flux distributions
from Schaerer \& de Koter (1997), plus Schmutz et al. (1992) WR models,
present between 2.8--5.2\,Myr (causing the hard EUV flux distribution). Ionization edges 
are indicated.}
\label{fig5}
\end{figure}

\subsection{Wolf-Rayet stars as the source of [O\,{\sc iv}] emission
in starbursts?}

Lutz et al. (1998) have recently proposed the identification of [O\,{\sc iv}]
25.91$\mu$m emission in several young starburst galaxies, including NGC\,5253.
They suggest that WR stars may provide the hard ionizing flux necessary
for such a highly excited feature. Our ISO spectrum suggests an upper 
limit for the [O\,{\sc iv}] line of 0.1$\times$10$^{-16}$ W m$^{-2}$
or [O\,{\sc iv}]/Br$\alpha$$\le$0.07, reasonably assuming any 
emission is restricted to the nuclear region. (Lutz et al. 
report 0.65 $\times$10$^{-16}$ W m$^{-2}$ based on higher
S/N data sets, implying [O\,{\sc iv}]/Br$\alpha$$\sim$0.35). 
Our observations indicate that [Ne\,{\sc v}] 14.3$\mu$m is not present 
in NGC\,5253.

Line blanketing softens the flux from the majority of early-type 
WR stars, such that negligible nebular [O\,{\sc iv}] 25.9$\mu$m emission is 
predicted from photoionization models of such stars, in 
contrast with early O stars (Table~\ref{table4}). 
This is the precise opposite of predictions using earlier O and WR models 
(Kurucz 1991; Schmutz et al. 1992)! Within the high ionization parameter 
nucleus of NGC\,5253, the low metallicity grids of Schaerer \& de Koter 
(1997) predict that observable nebular O\,{\sc iv} could be produced by 
early O dwarfs: [O\,{\sc iv}]/Br$\alpha$$\sim$0.5 for O3\,V stars at 
$\log~Q\sim$8.5. 

However, certain WR stars emit large numbers of photons with energies 
above the He\,{\sc ii} $\lambda$228 edge, necessary for [O\,{\sc iv}] 
emission. We include results of photo-ionization modelling of the 
WNE-w flux distribution in Table~\ref{table4}. From the table, 
[O\,{\sc iv}] 25.9$\mu$m emission is predicted at all ionization 
parameters, two orders of magnitude greater than O3\,V stars at 
parameters appropriate for the nucleus of NGC\,5253! Other high 
ionization IR nebular lines are predicted, including 
[Ne\,{\sc v}] 14.3$\mu$m ([Ne\,{\sc v}]/Br$\alpha$$\sim$6 at $\log~Q$=8.5). 

Since such emission features are not prominent in our ISO 
observations, we can exclude the possibility of
a significant number of such stars being present in NGC\,5253.
This is expected, since WNE-w stars are very rare in our Galaxy and the 
Magellanic Clouds. They have unusually low emission line fluxes\footnote{WR46 
has a He\,{\sc ii} $\lambda$4686 emission line flux of 
1.2$\times$10$^{35}$ erg\,s$^{-1}$ which is a factor of 4 lower than 
typical WNE stars (Schaerer \& Vacca 1998)}. Consequently, their 
identification would prove difficult in external galaxies from optical 
spectroscopic techniques.

Our results suggest that WR stars will only play a major role in the overall 
energy distribution of a young starburst region if they are very 
hot, and possess stellar winds that are relatively transparent. In the nearby
universe such stars are rare, so it is unlikely that they represent the 
`Warmers' responsible for 
the high IR excitation features observed in Active Galactic Nuclei 
(Terlevich \& Melnick 1985). However, metallicity also complicates matters 
yet further, affecting both the degree of EUV-line blanketing and the 
strength of the stellar wind (although a metallicity dependence  of 
mass-loss in WR stars has yet to be established observationally). 
If such a dependence is identified, weaker stellar winds {\it would} yield 
harder EUV flux distributions. Such a result would have an important 
effect on the possibility of WR stars providing the hard ionizing flux 
in very metal-poor starbursts, such as I~Zw~18 (de Mello  et al. 1998). 
More definitive answers await the calculation of a grid of line blanketed 
WR models for various combinations of mass-loss rates and metallicity. 
Crowther (1999) discusses this aspect in greater detail.

% 13-Jul-98 - Corrected for bug on Schaerer synthesis code
\begin{table} 
\caption[]{{\bf (a)} Results from photo-ionization models of evolutionary
synthesis calculations for an instantaneous
burst at 0.2$Z_{\odot}$ (Schaerer \& Vacca 1998) that are 
consistent with selected diagnostic IR line ratios for the 
\underline{nucleus} of NGC\,5253. Our preferred solutions 
(indicated by $\bullet$) are prior to the hot WR phase (2.9--4.7\,Myr) 
because of the uncertain WR flux distributions assumed in the models.}
\label{table5a} 
\begin{center} 
\begin{tabular}{
l@{\hspace{3mm}}
l@{\hspace{3mm}}
r@{\hspace{3mm}}
r@{\hspace{3mm}}
r@{\hspace{3mm}}
r@{\hspace{3mm}}
r@{\hspace{3mm}}
r@{\hspace{3mm}}
r} 
\hline 
Age            & log~$Q$  & \underline{Ar\,{\sc iii}} & \underline{S\,{\sc iv}}& \underline{Ne\,{\sc iii}}&\underline{S\,{\sc iv}} &\underline{Ne\,{\sc iii}}&\underline{S\,{\sc iv}} &\\
Myr            &          & S\,{\sc iv}               & Br$\alpha$             & S\,{\sc iv}              &Ne\,{\sc ii}              &Ne\,{\sc ii}             & S\,{\sc iii} &\\
\hline 
\multicolumn{2}{c}{Observed}& 0.09                    & $\sim$25               &$\le$1.4                       & $>$39                & $>$46?                & $>$1.4 &\\
\hline
\noalign{\smallskip}
2.0             & 8.0     &   0.10                    & 10                     & 1.0                      & 14                     & 30                      & 1.7 &\\
2.0             & 8.25    &   0.07                    & 14                     & 0.7                      & 60                     & 45                      & 2.4 &$\bullet$\\
2.0             & 8.5     &   0.08                    & 18                     & 0.6                      & 102                    & 63                      & 3.3 &\\
\noalign{\smallskip}
2.5             & 8.0     &   0.11                    &  9                     & 1.1                      & 21                     & 24                      & 1.3 &\\
2.5             & 8.25    &   0.08                    & 12                     & 0.8                      & 43                     & 35                      & 2.0 &$\bullet$\\
2.5             & 8.5     &   0.06                    & 16                     & 0.7                      & 79                     & 52                      & 2.8 &\\
\noalign{\smallskip}
3.0             & 8.0     &   0.09                    & 10                     & 1.0                      & 34                     & 32                      & 1.7 &\\
3.0             & 8.25    &   0.07                    & 14                     & 0.7                      & 66                     & 48                      & 2.5 &$\bullet$\\
3.0             & 8.5     &   0.05                    & 18                     & 0.6                      &117                     & 71                      & 3.5 &\\
\noalign{\smallskip}
4.0             & 8.0     &   0.10                    & 10                     & 1.0                      & 30                     & 30                      & 1.6 &\\
4.0             & 8.25    &   0.07                    & 13                     & 0.8                      & 59                     & 45                      & 2.3 &\\
4.0             & 8.5     &   0.05                    & 16                     & 0.6                      &100                     & 63                      & 3.0 &\\
\noalign{\smallskip}
4.5             & 8.25    &   0.09                    & 10                     & 0.9                      & 35                     & 32                      & 1.7 &\\ 
4.5             & 8.5     &   0.08                    & 11                     & 0.8                      & 68                     & 33                      & 2.0 &         \\
4.5             & 8.75    &   0.05                    & 17                     & 0.6                      &112                     & 68                      & 3.4 &\\ 
%\noalign{\smallskip}
%4.8             & 9.0     &   0.09                    & 11                     & 0.8                      &  32                    & 26                      & 1.9 &\\
\hline
\end{tabular}
\end{center}
\end{table}

\addtocounter{table}{-1}

\begin{table} 
\caption[]{{\bf (b)} Identical to Table~5a, except for  the \underline{core} 
of NGC\,5253.}
\label{table5b} 
\begin{center} 
\begin{tabular}{
l@{\hspace{3mm}}
l@{\hspace{3mm}}
c@{\hspace{3mm}}
r@{\hspace{3mm}}
c@{\hspace{3mm}}
r@{\hspace{3mm}}
r@{\hspace{3mm}}
r} 
\hline 
Age            & log~$Q$  & \underline{Ar\,{\sc iii}} & \underline{Ne\,{\sc ii}}& \underline{Si\,{\sc ii}}&\underline{S\,{\sc iii}}&\underline{S\,{\sc iv}}&\\
Myr            &          & Ne\,{\sc ii}               & Br$\alpha$             & Ne\,{\sc ii}              &Ne\,{\sc ii}             & Ne\,{\sc ii} &\\
\hline 
\multicolumn{2}{c}{Observed}& 0.08--0.3               & $\sim$3.3             & 1.2--2.2                & $\le$2                  & $<$0.35 &\\
\hline
 4.6            & 6.25       &   0.18                    & 3.5                   & 1.6                      & 1.0                      & 0.02                      &\\
%4.6            & 6.5        &   0.23                    & 3.1                   & 1.6                      & 1.6                      & 0.05                      &\\
\noalign{\smallskip}
4.7            & 6.0        &   0.12                    & 4.2                  & 1.7                      & 0.4                      & 0.01                     &\\
4.7            & 6.5        &   0.18                    & 3.2                   & 1.5                      & 0.9                      & 0.03                      &\\
4.7            & 7.0        &   0.29                    & 2.4                   & 1.4                      & 1.9                      & 0.19                      &\\
\noalign{\smallskip}
5.0            & 6.75       &   0.10                    & 3.3                   & 1.1                      & 0.9                      & 0.02                      &\\
5.0            & 7.5        &   0.14                    & 2.9                   & 0.7                      & 1.8                      & 0.13                      &\\
5.0            & 8.25       &   0.27                    & 2.1                   & 0.4                      & 3.1                      & 0.83                      &\\
\noalign{\smallskip}
5.2            & 7.5        &   0.08                    & 3.6                   & 0.5                      & 1.3                      & 0.05                      &\\
5.2            & 8.0        &   0.11                    & 3.7                   & 0.3                      & 1.7                      & 0.15                      &\\
5.2            & 8.5        &   0.19                    & 2.7                   & 0.2                      & 2.4                      & 0.55                      &\\
\noalign{\smallskip}
%8.0            & 6.25       &   0.01                    & 3.0                   & 1.5                      & 0.3                      & 0.00                      &\\
\hline
\end{tabular}
\end{center}
\end{table}

\section{Photo-ionization predictions for a burst of star formation}\label{sect7}

Our photo-ionization modelling has used  a single O or WR
star flux distribution, adjusted to differing ionization parameters. 
In reality, the nucleus and core of NGC\,5253 will contain stars spanning 
an enormous range
of masses and temperatures, even if they were formed in a single burst
of star formation. So, how valid is our assumption that the most massive
(and consequently the highest temperature) stars dominate the IR nebular
line fluxes? To investigate this, we have kindly been provided with 
evolutionary synthesis models (v 2.32) from Schaerer \& Vacca (1998). 
These utilise the low metallicity (0.2$Z_{\odot}$) Schaerer \&  de Koter (1997) 
flux distributions
for O-type stars, a Salpeter IMF, a mass range of 0.8--120$M_{\odot}$ and were 
calculated for differing ages according to the Meynet et al. (1994) 
evolutionary tracks after an instant burst of star formation. 

\begin{figure*}
\vspace{14.4cm}
\includegraphics{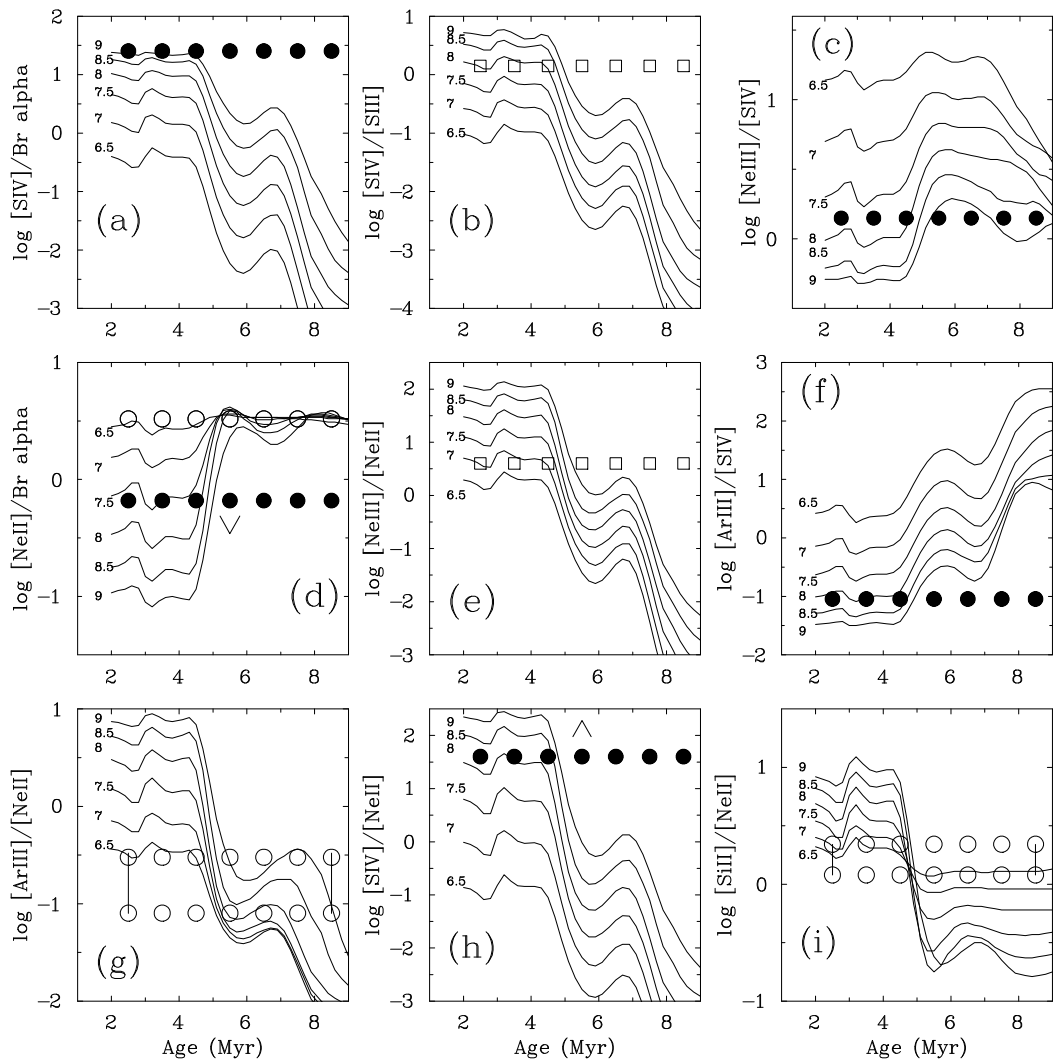}
\caption{Sensitivity of selected IR line ratios to ionization
parameter ($Q$) and age (Myr) for the instantaneous burst models 
from Schaerer \& Vacca (1998) at Z=0.2$Z_{\odot}$. Symbols are as 
in Figure~4.}
\label{fig6}
\end{figure*}

{}Fig.~\ref{fig5} presents EUV flux distributions produced by these models
for different ages. Excess high energy photons at $\lambda \le$228\AA\ 
are produced by WN and WC stars between 2.8--5.2\,Myr, since 
Schaerer \& Vacca (1998) utilised the low wind density Schmutz et al. 
(1992) WR model distributions. From  Section~\ref{sect6} such EUV excesses 
may be error. Nevertheless, detailed comparisons may be 
made with our predictions based on single O star models, treating
specific results during the WR phase with caution. 
We have generated a grid of 
photo-ionization models for ages in the range 2--10\,Myr and 
calculated fluxes for a fixed density of 10\,cm$^{-3}$,
with the total luminosity running from 
10$^{1}$, 10$^{2}$, ..., 10$^{11}$$L_{\odot}$, which covers a 
range in ionization parameter similar to our earlier calculations. 
The sensitivity  of IR diagnostic line ratios to ionization 
parameter and age (in Myr) are presented in Fig.~\ref{fig6},
where we again include observed line ratios in the nucleus (filled circles),
core (open circles) and combined core--nucleus region (open squares).

In Table~\ref{table5a}(a) we present selected age and ionization parameter
solutions that are in reasonable agreement with our IR diagnostics in 
the nucleus. The degree of consistency between alternative 
diagnostics is comparable to those from individual O star flux 
distributions (Table~\ref{table3a}[a]). 
Our earlier results for the nuclear region ionization parameter 
are supported, with $\log~Q$$\sim$8--8.5. The precise age is difficult 
to tightly constrain, but ages in the range 2--4\,Myr are in reasonable 
agreement with the observed line ratios. Recalling that the synthesis 
calculations between 2.9--4.7\,Myr are strongly affected by the 
presence of hot WR stars, our preferred solution is 2--3\,Myr.

{}For the core, Table~\ref{table5a}(b) presents results of our
photo-ionization models that are in reasonable agreement with our 
chosen mid-IR diagnostics, though the level of agreement is no
better than that obtained from individual O star flux distributions 
(Table~\ref{table3b}[b]). Once again, we find that the 
core ionization parameter remains uncertain, with log~Q=6.5 to 8 providing 
reasonable agreement for a fairly narrow range of ages, between 4.7 
to 5.2\,Myr. The only Wolf-Rayet stars predicted during this age range
are late WN stars, with a fairly soft EUV distribution, so that the 
evolutionary synthesis models should be appropriate. A significantly 
older population (5--10\,Myr) would require a very low ionization parameter 
(log~Q$\sim$6). As previously indicated, the entire volume of the core 
would only just be sufficient to contain the H\,{\sc ii} regions associated 
with 
2,500 O7\,V equivalents for this Q value, so we can exclude such a population.

Overall, our results from earlier calculations are broadly supported here,
based on more sophisticated flux distributions. Consequently, an important
result from this work is that meaningful results  may be obtained for 
very young  starburst regions based on individual O star flux 
distributions alone.

\section{Comparison with previous determinations}\label{sect8}

We now compare our present results for the hot, massive stellar content 
in NGC\,5253 with those obtained previously from ground 
(Aitken et al. 1982; Beck et al. 1996) and space-based (Lutz et al. 1996) IR 
observations, as summarised in Table~\ref{table6}. (We are unaware 
of any studies based on optical data sets). 

All previous mid-IR analyses relied 
on: (i) O-star flux distributions which did not account for non-LTE 
effects or  stellar winds, or in many cases the low metallicity 
of the galaxy; (ii) inappropriate line ratios (given the results from 
ground- and space- based observations), and (iii) assumed elemental 
abundances (often solar composition) and ionization parameter. 
Consequently, previous approaches may lead to imprecise results, although 
all studies support the presence of hot, young O-type stars in 
its central starburst, with characteristic temperatures 
ranging from 40kK to $\ge$50kK.

\begin{table*} 
\caption[]{Summary of comparison between stellar temperatures 
and ionization parameters derived for NGC\,5253 using IR diagnostics.
Theoretical O star fluxes are from Hummer \& Mihalas (1970, HM70), 
Kurucz (1991, K91) and Schaerer \& de Koter
(1997, SdK97), while photo-ionization models employed are {\sc mappings ii}
(Sutherland \& Dopita 1993, SD93) and {\sc cloudy} (Ferland 1996, F96).}
\label{table6} 
\begin{center} 
\begin{tabular}{lllclcrrcl} 
\hline 
Study& IR Diagnostics& \multicolumn{2}{c}{-- O star Fluxes --} & Photo- & Abund. & 
$T_{\rm eff}$ & log $Q$ & Age & Notes\\
     &               &  Ref. & $Z_{\odot}$ & Code & $Z_{\odot}$ & kK &  & Myr &\\
\hline
Aitken et al. 1982   & [S\,{\sc iv}] & HM70 & 1.0 & --  & 1.0 & $\ge$50 & -- & &\\
Beck et al. 1996     & [S\,{\sc iv}], [Ar\,{\sc iii}], [Ne\,{\sc ii}] 
                     & K91 & 0.1--1.0 & SD93 &1.0 & 40--45 & -- & -- & \\
Lutz et al. 1996     & [Ne\,{\sc ii-iii}] & K91 & 1.0 & F96 & 1.0 & 48.5 & 8.0 & -- & $Q$ assumed \\
This work (nucleus)  & [S\,{\sc iv}], [Ar\,{\sc iii}], [Ne\,{\sc iii}], Br$\alpha$ & SdK97
                     & 0.2 & F96 & 0.25 & 38--50 & $\sim$8.25 & 2--3 &  \\
This work (core)     & [Ne\,{\sc ii}], [Ar\,{\sc iii}], [S\,{\sc iii}], Br$\alpha$  & SdK97 & 0.2 & F96 & 0.25 & 
                     $\sim$35 &$\le$8 & 4.7--5.2 & \\
\hline
\end{tabular}
\end{center}
\end{table*}

Our comparison with previous results well illustrates the effect 
of different O-star flux distributions and elemental abundances, even 
with an identical photo-ionization code and IR line diagnostics being 
employed. For example, Lutz et al. (1996) obtained a characteristic O star
temperature of 48.5kK from the  observed  [Ne\,{\sc iii}]/[Ne\,{\sc ii}] 
ratio  since they adopted {\it solar} Kurucz (1991) models, {\it solar}
elemental abundances and log~$Q$=8. From our analysis, neglecting the 
spatial information for line formation in NGC\,5253 and adopting 
log~$Q$=8 implies a stellar temperature of 36kK for the combined 
core--nucleus region using [Ne\,{\sc iii}]/[Ne\,{\sc ii}] from 
Fig.~\ref{fig3}(e). Alternatively, a stellar temperature of 39kK  would 
result from the 
observed [S\,{\sc iv}]/[S\,{\sc iii}] ratio with an identical ionization 
parameter (Fig~\ref{fig3}[b]). 

Having established constraints on the stellar content of the nucleus and
core of NGC\,5253 we are able to compare ages obtained from the
derived stellar temperatures with those from evolutionary synthesis
calculations. A minimum characteristic stellar temperature of 
$\ge$38kK for the nucleus implies a minimum characteristic mass of 
$\sim$35$M_{\odot}$.  Using the result of Kunze et al. (1996) 
that the upper mass cut-off of a cluster is about 5$M_\odot$ greater than
the single-star equivalent mass gives a lower limit to the mass cutoff of 
40$M_\odot$ for the hot nucleus, with an age no greater than around 
3\,Myr (Meynet et al. 1994). This is in reasonable agreement with 
the evolutionary synthesis calculations for log~$Q$$\sim$8, and 
is typical of recent age estimates for the central nucleus 
of NGC\,5253 from the WR population (Schaerer et al. 1997). 
(For $T_{\rm eff}$$\approx$46kK the implied hottest stellar type is O4 or O5
with mass $\sim$60$M_{\odot}$ at a corresponding age of $\sim$2.5\,Myr).

For the core region, a stellar temperature of $\sim$35kK corresponds to 
the hottest spectral type being equal to or later
than O8V or O9V (Table~\ref{table2}), and a equivalent mass of 
$\sim$25$M_{\odot}$. This gives an upper mass cutoff of 30$M_\odot$, 
indicating an age of around 4--5\,Myr, in good agreement with that 
implied from our synthesis calculations. 
However, we note that our  
determination is based on the entire core of NGC\,5253, spanning 
several hundred parsec. We do not claim that the massive stellar 
population in the core of  NGC\,5253 was formed coevally $\sim$5\,Myr ago.  
Our results are almost certainly biased towards the youngest, most massive 
clusters exterior to the central nucleus. In reality, the content of the core
will span a large range of ages. The weakness of [Ar\,{\sc ii}] emission
does appear to rule out a particularly old population. Vanzi \& 
Rieke (1997) concluded from near-IR 
spectroscopy, that NGC\,5253 is young ($\sim$8Myr) and contains high 
temperature stars ($\ge$40kK) since He\,{\sc i} 1.70$\mu$m is present, 
[Fe\,{\sc ii}] emission (from SN remnants) is very weak, as 
are the CO absorption bands (produced by red giants). Davies et al. 
(1998) estimated a greater age of $\sim$10--12\,Myr for regions 
$\sim$7\,arcsec from the central nucleus. 

Again, neglecting the spatial origin of low and high excitation lines
and assuming a uniform ionization parameter of $\log~Q$=8, an age 
of $\sim$4.8\,Myr is obtained from [Ne\,{\sc iii}]/[Ne\,{\sc ii}] following
Fig.~\ref{fig6}(e). Alternatively, ages of either 2.9 or 4.4\,Myr are implied 
from  the observed [S\,{\sc iv}]/[S\,{\sc iii}] ratio (Fig.~\ref{fig6}[b]).

To summarize, many models and techniques may be
applied to IR observations of this galaxy. While they differ 
in details they agree with each another in the general picture of NGC\,5253. 
This galaxy has a very young starburst in the nucleus and a slightly older 
one within the core. It is remarkable for the youth of the starburst, probably 
no more than 3\,Myr, and for the very massive stars it formed. (For stars to
reach the WC phase in such a low metallicity environment requires very
massive progenitors of 60--120$M_{\odot}$ (Maeder \& Meynet 1994).)
 
\section {Discussion and Conclusions}\label{sect9}

NGC\,5253 is widely acknowledged to be an 
excellent, in some respects unequalled, laboratory for the study of 
intense starbursts. This is because it is relatively close, contains
a very young starburst, and because its central region is so varied,
containing many sub-regions and clusters of different ages and stellar
content. (The patchiness of the central region of NGC\,5253 is probably 
related to its extreme youth; there has not been time for stars to 
migrate away from their points of formation and for the original structure 
of star formation to be lost).  NGC\,5253 also has the special advantage 
of dwarf galaxies: in contrast with large spiral galaxies bursts are seen
against a very low intra-burst star formation rate rather than  steady 
star formation activity.

It is therefore rather troubling that the properties of NGC\,5253 at which
we arrive in this paper depend so crucially on the 
high spatial resolution ground-based data. The characteristics of NGC\,5253
that make it so interesting also make it very hard to interpret
observations that could not, for example, distinguish between the 
nucleus and core. 
If we had not been able to assign the different infrared lines to their
appropriate sources, we would have found highly contradictory line 
ratios and an imprecise effective stellar temperature of the ionizing O
stars. Even with the existing ground- and space- based observations, some 
important results await confirmation; e.g. What fraction of the 
[S\,{\sc iii}] nebular flux is located in the nucleus?

Nevertheless, it is clear from our results that the nucleus of NGC\,5253
contains extremely young, massive, hot stars, with a high characteristic 
ionization parameter. The surrounding core contains cooler stars, 
with a lower ionization
parameter, and undoubtedly contains stars spanning a wide range of ages. The
nucleus of NGC\,5253 is not 
unique in this respect; the stellar content of 30 Doradus in the LMC, for 
example, shares many of its attributes, and deserves greater study at 
mid-IR wavelengths. One important result of our study is that
estimates of the global properties of young starbursts from the 
effective temperatures of their most massive constituents appears 
to be a reasonable approach. 

Unfortunately it is not yet possible to readily compare the spatial 
density of O stars in the  super-star nucleus of NGC\,5253 with other
giant H\,{\sc ii} regions. Definitive answers await high spatial resolution 
IR observations, but the ionization parameter for the nucleus indicates 
a dense concentration. At the very least, 1\,000 O7\,V ``equivalents'' occupy 
a region of radius $\le$15 parsec. For comparison, 
$\sim$100 early O stars, with a {\it similar} total Lyman ionizing flux are 
located within a radius of 10 parsec from R136a in 30~Doradus  
(Massey \& Hunter 1998; Crowther \& Dessart 1998). 

Our principal result illustrates the challenge faced with nearly all 
observations of galaxies; many different sources fall into a single
aperture. This is particularly acute for dwarf starburst galaxies, where
there is less background of star formation, where the smoothing effects of
shear  and spiral arm forces are absent, and where the starburst activity
is likely to be concentrated in one or a few clumps a few parsecs in 
size. Consequently, high spatial resolution observations
of NGC\,5253 and other nearby starbursts are urgently sought at near-IR 
and mid-IR wavelengths, and quantitative results for distant 
starbursts should, for the moment, be treated with caution.

\section{Acknowledgements}
This work is based on observations with ISO, an ESA 
project with instruments funded by ESA Member States (especially the PI 
countries: France, Germany, the Netherlands and the United Kingdom) with 
the participation of ISAS and NASA. We thank John Hillier for 
providing his stellar atmosphere code, Daniel Schaerer for providing 
us with his grid of evolutionary synthesis calculations, and appreciate 
useful discussions with Gary Ferland and  Jean-Ren\'{e} Roy. An anonymous
referee provided extremely useful comments on this work. PAC 
is a Royal Society University Research Fellow. SCB was supported 
by the US-Israel Binational Science Foundation grant 
94-00303. PSC appreciates hospitality at University College London 
where the initial proposal for this work was first begun.
This research  has made use of the NADA/IPAC Extragalactic Database (NED) 
which is operated by the Jet Propulsion Laboratory, California Institute of 
Technology, under contract with the National Aeronautics and Space 
Administration.

\label{lastpage}

\end{document}